\documentclass[12 pt]{article}
\usepackage{a41}
\usepackage{cite}
\usepackage{color}
\usepackage{epsfig}
\usepackage{amssymb}
\usepackage{amsmath}
\usepackage[T1]{fontenc}
\usepackage[latin1]{inputenc}

\def \f14{\hbox{\large$\frac{1}{4}$}}
\def \f12{\hbox{\large$\frac{1}{2}$}}

\newcommand{\KI}{ {\mathcal{K}} }

\newcommand{\twz}{ {\text{tw2}} }

\newcommand {\PP}{{\mathbb{P}}}

\newcommand {\kln}[1]{\left( #1 \right)}

\newcommand {\kls}[1]{\left\{ #1 \right\}}

\newcommand {\kle}[1]{\left[ #1 \right]}

\newcommand {\Matel}[3]{\bigl< #1 \bigl|\; #2\, \bigr| #3 \bigr>}

\newcommand {\im}{{\text{i}}}
\newcommand {\e}{{\text{e}}}

\newcommand\pvec{\mbox{\boldmath $p$}}
\newcommand\qvec{\mbox{\boldmath $q$}}
\newcommand\rvec{\mbox{\boldmath $r$}}
\newcommand\lvec{\mbox{\boldmath $l$}}

\begin{document}
\noindent
\sloppy
\thispagestyle{empty}
\begin{flushleft}
{DESY 08-091} 
\\
SFB-CPP-08/101\\
December 2008
\end{flushleft}

\vspace*{\fill}
\begin{center}

{\LARGE\bf \boldmath Target mass and finite $t$ corrections }

\vspace{3mm}
{\LARGE\bf to diffractive deeply inelastic scattering}

\vspace{2cm}
\large Johannes Bl\"umlein$^a$,  Dieter Robaschik$^{a,b}$ and Bodo Geyer$^c$

\vspace{2em}
\normalsize {\it $^a$~Deutsches Elektronen--Synchrotron, DESY,\\ Platanenallee
6, D--15738 Zeuthen, Germany} \\

\vspace{2em} {\it $^b$~Brandenburgische Technische Universit\"at Cottbus,
Fakult\"at 1,}\\ {\it PF 101344, D--03013 Cottbus, Germany} \\

\vspace{2em} $^c${\it Center for Theoretical Studies and
Institute of Theoretical Physics,\\ Leipzig University, Augustusplatz~10, D-04109~Leipzig,
Germany} \\

\vspace{2em}
\end{center}

\vspace*{\fill}
\begin{abstract}
\noindent
The quantum field theoretic treatment of inclusive deep--inelastic 
diffractive scattering given in a previous paper \cite{BGR2006} is 
discussed in detail using an equivalent formulation with the aim to 
derive a representation suitable for data analysis. We consider the 
off-cone twist--2 light-cone operators to derive the target mass and 
finite $t$ corrections to diffractive deep--inelastic scattering and
deep--inelastic scattering. The corrections turn out to be at most 
proportional to $x |t|/Q^2,  x M^2/Q^2,~~x = x_{\rm BJ}~{\rm or}~~x_\PP$,
which suggests an expansion in these parameters. Their contribution 
varies in size considering diffractive scattering or meson--exchange 
processes. Relations between different kinematic amplitudes which are 
determined by one and the same diffractive GPD or its moments are derived. 
In the limit $t, M^2 \rightarrow 0$ one obtains the results of 
\cite{Blumlein:2001xf} and \cite{Blumlein:2002fw}. 
\end{abstract}

\vspace{1cm}
{\small
\noindent PACS: 24.85.+p, 13.88.+e, 11.30.Cp\\ Keywords: Diffractive Scattering, Target
Mass Effects, Finite momentum transfer corrections, Twist decomposition, Nonlocal light-cone
operators, Multivalued distribution amplitude,
Generalized Bjorken limit.}

\section{Introduction} \renewcommand{\theequation}{\thesection.\arabic{equation}}
\setcounter{equation}{0}

\vspace{1mm}
\noindent
The process of deep--inelastic diffractive lepton--nucleon scattering
can be measured at high energy
colliders and constitutes a large fraction of the inclusive statistics, 
although being a 
semi--inclusive process. It was first observed at the electron--proton collider HERA some 
years ago~\cite{Derrick:1993xh} and is now  measured in detail~\cite{Derrick:1996ma}. 
The structure function $F_2^D(x,Q^2)$ was extracted. In the same manner it is desirable to compare  
the longitudinal
diffractive structure function $F_L^D(x,Q^2)$  with the 
longitudinal structure function in the inclusive case~\cite{Adloff:1996yz,FLH1}. The 
measurement 
of the polarized diffractive structure functions $g_{1,2}^D(x,Q^2)$ will be possible at 
future facilities like EIC \cite{EIC}, which are currently planned.
The experimental measurements clearly showed that the scaling violations of the
deep-inelastic and the diffractive structure functions in the deep-inelastic regime,
after an appropriate change of kinematic variables, are the same. Furthermore, the ratio
of the two quantities, did not vary strongly, cf. \cite{Abramowicz:1996ha}. While the
former property is clearly of perturbative nature, the latter is of non--perturbative
origin. For diffractive scattering, however, another mass scale is of importance, which is 
given by the invariant mass $t = (p_2 - p_1)^2$. Here $p_{1(2)}$ denote the 4--momenta of 
the incoming and outgoing proton, where for the latter a sufficiently large rapidity gap 
between this particle and the remainder final state hadrons is demanded as process 
signature. A similar class of processes are the so-called meson--exchange processes, cf. 
e.g. \cite{Kopeliovich:1996iw}, where
the finite rapidity gap is not required, but the r\^{o}le of the formerly diffractive final 
state proton is taken by a leading hadron, which distinguishes itself due to its high 
momentum from the remaining hadrons. Also in this case one 
may try a leading twist description, although the signature for this process is 
less clear than in the diffractive case. 

The process of deep--inelastic diffractive scattering was first described
phenomenologically \cite{PHEN}. A consistent field-theoretic description of the process
requires factorization for the twist--2 contributions \cite{Berera:1995fj}.
It is due to this description that reference to specific pomeron models are 
thoroughly avoided. In Refs.~\cite{Blumlein:2001xf,Blumlein:2002fw,Blumlein:2002ax} 
two of the present authors gave a corresponding field--theoretic description of the 
process in the limit $t, M^2 \rightarrow 0$. In \cite{Blumlein:2001xf} 
we proved that under these conditions the scaling violations 
for diffractive scattering and inclusive deeply inelastic scattering are the same, up to 
a change in the momentum-fraction variable in the former case.

At low 4--momentum transfer $Q^2$ both target mass $(M^2)$ and finite 
momentum transfer 
$(t)$ corrections have to be considered for the diffractive and leading hadron processes 
with meson exchange. In the deep-inelastic case the target mass corrections
were studied in Refs.~\cite{Nachtmann:1973mr, Georgi:1976ve,Blumlein:1998nv,Piccione:1997zh}, 
see also \cite{TM2}. The kinematics of the diffractive and leading hadron
processes 
is similar to that in deeply-virtual non--forward scattering. Considering this general 
class of processes,
one finds that the treatment of target mass effects and finite $t$--effects can 
only be performed by combining both, see~\cite{gey2,belm}. If compared with the 
deep--inelastic case the number of hadronic structure functions enlarges in the 
diffractive case from two to four for  unpolarized scattering and to eight for  polarized 
scattering, as shown in \cite{Blumlein:2001xf,Blumlein:2002fw}, if the 
general kinematics is considered. In Ref.~\cite{BGR2006} we worked out these corrections 
for the hadronic tensor in general, yet without quantifying the result. If one departs 
form the limit $t, M^2 \rightarrow 0$ the corresponding representations require to carry 
out a one-dimensional {\sf definite} integral which kinematically relates the two proton 
momenta $p_1$ and $p_2$. As the integration is to be performed over  unknown 
non-perturbative functions there is no a priori experimental way to unfold the 
non-perturbative  distributions, which also would invalidate the partonic description 
in case of diffractive scattering. Moreover, the $M^2$ and $t$ effects dealt with in this 
case are not yet complete, since there emerge other contributions more in the scattering 
cross section. One may expand the complete solution in two variables 
$t/Q^2, M^2/Q^2$. It is found that these terms multiply at least with a factor $x = x_{{\rm BJ}(\PP)}$, 
which is bounded in the diffractive case to values below 0.01 and in the meson--exchange case 
$\lessapprox 0.3$. Thus the leading terms beyond $t, M^2 = 0$ give a good first estimate for 
the corrections. The further corrections turn out to be widely suppressed in the 
diffractive case, while they are larger for leading particle cross sections in the 
meson--exchange case. 

In the present paper we will discuss both the unpolarized and polarized case. 
The paper is organized as follows. In Section~2
we derive the differential scattering cross section for inclusive diffractive scattering 
at the Lorentz level. Main aspects of the relation of this 
process to the Compton 
amplitude within the light--cone expansion including finite $M^2$ and $t$ effects are 
summarized in Section~3. The hadronic tensors for the unpolarized and polarized case 
are expanded in terms of the variables $t/Q^2,~~M^2/Q^2$ in Section~4 to show the size of 
the correction terms. Section~5 contains 
the conclusions. In Appendix~A we summarize some kinematic relations. The present formalism
is specified to the case of deep--inelastic forward scattering (DIS) in 
Appendix~B, where we obtain the target mass corrections given in 
\cite{Georgi:1976ve,Blumlein:1998nv,Piccione:1997zh}  before.
\section{The Lorentz Structure}
\renewcommand{\theequation}{\thesection.\arabic{equation}}
\setcounter{equation}{0}

\vspace{1mm}
\noindent
The process of deep--inelastic diffractive scattering belongs to
the class of semi--inclusive processes. It is described by an effective $2 \rightarrow 3$ 
diagram, cf.~Figure~1 Ref.~\cite{Blumlein:2001xf}, with incoming and outgoing 
charged lepton and nucleon lines and an effective 4-vector for all the other hadron lines 
in the final state, which are well separated in rapidity from the outgoing diffractive nucleon 
line. 

The differential scattering cross section for single--photon exchange is given by
\begin{equation}
\label{eqD1}
 \text{d}^5 \! \sigma_{\rm diffr} = \frac{1}{2(s-M^2)} \, \frac{1}{4} \;
    dPS^{(3)} \sum_{\rm spins}\frac{e^4}{Q^2} \, L_{\mu\nu} W^{\mu\nu}~.
\end{equation}
Here $s=\kln{p_1 + l_1}^2$ is the cms energy squared of the process  and $M$ 
denotes
the nucleon mass. The phase space $d P S^{\kln{3}}$ depends on five variables
since the mass $M_X$ of the diffractively produced inclusive set of hadrons varies. We 
choose as  basic variables
\begin{eqnarray}
 x_{\rm BJ} &=& \frac{Q^2}{Q^2 + W^2 - M^2} = - \frac{q^2}{2 \, q p_1} \; , \\
 y          &=& \frac{Q^2}{x_{\rm BJ}(s-M^2)}~, 
\end{eqnarray}
$t=\kln{p_2 - p_1}^2$ the 4--momentum difference squared between incoming and outgoing
nucleon, a variable describing the non-forwardness w.r.t.~the incoming
proton direction,
\begin{equation}
\label{eqV1}
x_{\PP} = \frac{Q^2 + M_X^2 - t}{Q^2 + W^2 - M^2} = - \frac{q p_-}{q p_1}
\geq x_{\rm BJ}~, 
\end{equation}
and the angle $\phi_b $ between the lepton plane $\pvec_1 \times \lvec_1 $ and
the hadron plane  $\pvec_1 \times \pvec_2$,
\begin{equation}
\label{eqV3}
\cos(\phi_b) = \frac{(\pvec_1 \times \lvec_1).(\pvec_1 \times  \pvec_2)}
                 {|\pvec_1 \times \lvec_1 ||\pvec_1 \times  \pvec_2|}~.
\end{equation}
Here $Q^2 = - q^2$ denotes
the photon virtuality  and
$W$ is the hadronic mass with $W^2 = \kln{p_1 + q}^2$. We also refer to $x = Q^2/qp_+$.
It is useful to introduce the 4--vectors
\begin{equation}
p_\pm = p_2 \pm p_1~.
\end{equation}
The diffractive mass squared is given by
$M_X^2 = \kln{q - p_-}^2$.
The momenta $p_\pm$ obey
\begin{equation}
(p_+~p_-) = 0, \qquad \frac{p_+^2}{p_-^2} = \frac{4 M^2}{t} - 1\,.
\end{equation}
For later use we refer to the non-forwardness $\eta$ and the variable $\beta$ defined by
\begin{equation}
 \eta = \frac{q p_-}{q p_+} = \frac{-x_{\PP}}{2-x_{\PP}}
 \in \kle{-1 \, , \, \frac{-x}{2-x} } \; ,
\qquad
  \beta  =   \frac{q^2}{2\,qp_-} =\frac{x_{\rm BJ}}{x_{\PP}} \leq 1\,.
\end{equation}
The variable $x_{\PP}$ is directly related to $\eta$ but is more commonly used 
in 
experimental analyzes,
\begin{equation}
x_{\PP} = \frac{2 \eta}{\eta -1}~.
\end{equation}
More kinematic invariants are given in Appendix~A.

The transverse momentum variable, introduced as $\hat\pi_-$, \cite{BGR2006}, or $\pi_- = 
-\eta \hat\pi_- $   is of special importance, 
\begin{equation}
\label{tmv}
\pi_- = p_- - {p_+}{\eta}, \qquad (q \pi_-) = 0\,.
\end{equation}
Later on it plays the role of an expansion parameter.
The variables $x_{\rm BJ}, x_\PP, \beta$ and $\eta$ obey the inequalities
\begin{align}
&0 \leq x_{\rm BJ} \leq x_\PP \leq 1, \qquad 0 \leq x_{\rm BJ} \leq \beta \leq 1,
\\
-\infty \leq 1& - \frac{2}{x_{\rm BJ}} \leq 1 - \frac{2 \beta}{x_{\rm BJ}} = 
\frac{1}{\eta} \leq -1 \leq \eta
\leq \frac{-x_{\rm BJ}}{2-x_{\rm BJ}} \leq 0~.
\end{align}

For the spin averaged cross section, the leptonic tensor is symmetric.
Taking into account conservation of the electromagnetic current one obtains
\cite{Blumlein:2001xf}
\begin{eqnarray}
\label{eqD2}
 W_{\mu\nu}^s
&=&
  -g_{\mu\nu}^T  W_1^s +  p_{1\mu}^T p_{1\nu}^T \frac{W_2^s}{M^2}
   + p_{2\mu}^T  p_{2\nu}^T \frac{W_4^s}{M^2}
   + \kle{ p_{1\mu}^T  p_{2\nu}^T +  p_{2\mu}^T  p_{1\nu}^T } \frac{W_5^s}{M^2} \; .
\end{eqnarray}
Here and in the following we do not assume that azimuthal 
integrals are performed as sometimes is done in experiment. 
In the latter case the number of contributing structure function reduces.

In the case of polarized nucleons we consider the initial state spin--vector 
$S_1\equiv S,~S^2 = -M^2$,  only and sum over the spin of the outgoing 
hadrons. One usually refers to the longitudinal $(||)$ and transverse 
($\perp$) spin projections choosing
\begin{eqnarray}
S_{||}    &=& (\sqrt{E^2-M^2};0,0,0,E)~, \\
S_{\perp} &=& (0;\cos\gamma,\sin\gamma,0) M~,
\end{eqnarray}
in the laboratory frame with $p_1 = (E;0,0,\sqrt{E^2-M^2})$, with $S.p_1 = 0$.
Here $\gamma$ denotes the azimuthal angle. In the case of 
longitudinal polarization the contraction of $S_{||}$ with $l_1$ and $p_2$ being nearly
collinear to $p_1$ are of ${\cal O}(\mu^2/Q^2),~~\mu^2 = |t|, M^2$, see Appendix A.

The antisymmetric part of the hadronic tensor was derived in 
\cite{Blumlein:2002fw} and is given by
\begin{alignat}{10}
\label{eqH2}
W_{\mu\nu}^a &=&~~
i \left[ {p}_{1\mu}^T {p}_{2\nu}^T-
{p}_{1\nu}^T {p}_{2\mu}^T \right] \varepsilon_{p_1 p_2 q S}
&\frac{W_1^a}{M^6}&~~
                &+&~~
i \left[ {p}_{1\mu}^T \varepsilon_{\nu S p_1 q}
     - {p}_{1\nu}^T \varepsilon_{\mu S p_1 q} \right]
&\frac{W_2^a}{M^4}&
\nonumber \\ &+&~~
i \left[ {p}_{2\mu}^T \varepsilon_{\nu S p_1 q}
     - {p}_{2\nu}^T \varepsilon_{\mu S p_1 q} \right]
&\frac{W_3^a}{M^4}&~~
                &+&~~
i \left[ {p}_{1\mu}^T \varepsilon_{\nu S p_2 q}
     - {p}_{1\nu}^T \varepsilon_{\mu S p_2 q} \right]
&\frac{W_4^a}{M^4}&
\nonumber \\ &+&~~
i \left[ {p}_{2\mu}^T \varepsilon_{\nu S p_2 q}
     - {p}_{2\nu}^T \varepsilon_{\mu S p_2 q} \right]
&\frac{W_5^a}{M^4}&~~
                &+&~~
i \left[ {p}_{1\mu}^T {\varepsilon}_{\nu p_1 p_2 S}^T
     - {p}_{1\nu}^T {\varepsilon}_{\mu p_1 p_2 S}^T \right]
&\frac{W_6^a}{M^4}&
\nonumber\\ &+&~~
i \left[ {p}_{2\mu}^T {\varepsilon}_{\nu p_1 p_2 S}^T
     - {p}_{2\nu}^T {\varepsilon}_{\mu p_1 p_2 S}^T \right]
&\frac{W_7^a}{M^4}&~~
                &+&~~   i\; \varepsilon_{\mu \nu q S}
&\frac{W_8^a}{M^2}&~,
\end{alignat}
where $\varepsilon_{\mu\nu\alpha\beta}$ denotes the Levi--Civita symbol.
The kinematic factors above are constructed out of the
four--vectors $q,p_1,p_2$ and $S$ as well as $g_{\mu\nu}$
and $\varepsilon_{v_0 v_1 v_2 v_3}$ using
\begin{eqnarray}
\label{trans}
p_\mu^T &=& p_\mu - q_\mu \frac{q.p}{q^2},\,\,\,
  g_{\mu\nu}^T=  g_{\mu\nu} - \frac{q_\mu q_\nu}{q^2},          
\\
{\varepsilon}_{\mu v_1 v_2 v_3}^T            &=&
    {\varepsilon}_{\mu v_1 v_2 v_3}            -
    {\varepsilon}_{q v_1 v_2 v_3} \frac{q_\mu}{q^2}~,  \\
{\varepsilon}_{\mu \nu v_1 v_2}^{TT}            &=&
    {\varepsilon}_{\mu \nu v_1 v_2}            -
    {\varepsilon}_{q \nu v_1 v_2} \frac{q_\mu}{q^2}
  - {\varepsilon}_{\mu q v_1 v_2} \frac{q_\nu}{q^2}~.
\end{eqnarray}
One may rewrite (\ref{eqH2}) into an equivalent form using the 
Schouten-identities~\cite{SCHOUT}.

Target mass and finite $t$ corrections to the differential scattering cross
section (\ref{eqD1}) in the leading twist approximation emerge from three sources:~~$(i)$ from 
kinematic terms at the Lorentz level after contracting the leptonic and hadronic tensor; 
~~$(ii)$ from the expectation value of the Compton operator; ~~$(iii)$
the $t$-behaviour of the non--perturbative distribution functions.

We will first consider the contributions $(i)$ and discuss the terms $(ii)$ in Section~4.
The non-perturbative effects cannot be calculated by rigorous methods within Quantum
Chromodynamics at present, but are left to phenomenological models or are determined 
through fits to data, cf.~\cite{PHEN}. 

For pure photon exchange the leptonic tensor is given by
\begin{eqnarray}
L_{\mu\nu} = 2(l_{1\mu} l_{2\nu} + l_{2\mu} l_{1\nu} - g_{\mu\nu} l_1.l_2 - i 
\varepsilon_{\mu\nu\alpha\beta} l_1^\alpha q^\beta),
\end{eqnarray}
cf.~\cite{LR}, in case of longitudinal lepton polarization.

We consider the Bjorken limit, 
\begin{eqnarray}
2 p_1.q = 2 M \nu 
\rightarrow \infty,~~~~p_2.q \rightarrow \infty,~~~~Q^2 
\rightarrow \infty,~~{\rm with}~~~~x_{\rm BJ}~~~{\rm and}~~~x_\PP = {\sf 
fixed.}
\end{eqnarray}
Here, 
\begin{eqnarray}
M W_1^s   &\rightarrow& F_1 \\ 
\nu W_k^s &\rightarrow& F_k,~~~k=2,4,5~,  
\end{eqnarray}
with $\nu = y (s-M^2)/(2M)$. 

In the unpolarized case we obtain in the limit
$M^2, t \rightarrow 0$ w.r.t. the kinematics of the momenta $p_1$ and $p_2$, 
keeping the target mass dependence  
\begin{eqnarray}
\label{xs1}
\frac{d^s \sigma^{\rm unpol}}{d x_{\rm BJ}\, d Q^2} = \frac{2 \pi \alpha^2}{Q^4 x_{\rm BJ}} 
\left[2xF_1 \cdot y^2
                          + \left[F_2 + (1-x_\PP) F_4 
                          + (1-x_\PP)^2 F_5\right] 
\cdot 2\left(1-y - \frac{x^2_{\rm BJ} y^2 M^2}{Q^4}\right)\right], 
\end{eqnarray}
where $F_k = F_k(x_{\rm BJ},x_\PP,Q^2;t)$ are the diffractive 
structure functions, cf.~\cite{Blumlein:2001xf}. The correction terms are
of ${\cal O}(M^2/Q^2,~t/Q^2)$. In the limit $M^2, t \rightarrow 0$ the azimuthal 
dependence on $\phi_b$ vanishes.

Likewise we obtain in the polarized case for longitudinal nucleon 
polarization,
\begin{eqnarray}
\label{xs2}
\frac{d^3 \sigma^{\rm pol}(\lambda, \pm S_{||})}{d x_{\rm BJ} d Q^2 d x_\PP} &=& \mp 4 \pi s 
\lambda \frac{\alpha^2}{Q^4} \left[y \left(2-y - \frac{2x_{\rm BJ}yM^2}{s}\right) xg_1 
- 4x_{\rm BJ}y \frac{M^2}{s} g_2\right]~,\\
\label{xs3}
\frac{d^4 \sigma^{\rm pol}(\lambda, \pm S_{\perp})}{d x_{\rm BJ} d Q^2 d x_\PP. d \Phi} &=& \mp 4 
\pi s 
\lambda \sqrt{\frac{M^2}{s}} \frac{\alpha^2}{Q^2} \sqrt{x_{\rm BJ}y \Bigl[1-y- \frac{x_{\rm BJ}y 
M^2}{s}\Bigr]}
\cos(\gamma - \Phi) \left[yx_{\rm BJ} g_1 + 2 x_{\rm BJ} g_2\right]~.
\nonumber\\
\end{eqnarray}
Here $\phi$ denotes the angle between the $\vec{l}_1-\vec{S}$ and the $\vec{l}_1-\vec{l}_2$
plane and $\alpha$ is the anle between $\vec{l}_1$ and $\vec{S}$. 
The structure functions $g_{1,2}(x_{\rm BJ},x_\PP,Q^2;t)$ are obtained from $W_2^a, W_3^a, W_4^a,
W_5^a$ and $W_8^a$ by
\begin{eqnarray}
g_1 &=& \frac{p.q_1}{M^2} W_8^a~,
\\
g_2 &=& \frac{(p.q_1)^3}{q^2 M^4} \left[W_2^a + (1-x_\PP)[W_3^a + W_4^a] + (1-x_\PP)^2
W_5^a \right]
\end{eqnarray}
and the different structure functions $F_i$ and $g_i$ depend on the variables
$x_{\rm BJ},x_\PP,Q^2$ and $t$.

\section{The Compton Amplitude}
\renewcommand{\theequation}{\thesection.\arabic{equation}}
\setcounter{equation}{0}

\vspace{1mm}
\noindent
The hadronic tensor for deep--inelastic diffractive scattering can be obtained from a 
Compton amplitude as has been outlined in \cite{Blumlein:2001xf,Blumlein:2002fw,BGR2006} 
before. We limit the description to the level of the twist--2 contributions, where
factorization holds for the semi--inclusive diffractive process \cite{Berera:1995fj}.  
Furthermore, A.~Mueller's generalized optical theorem \cite{Mueller:1970fa}
allows to move the final state proton into an initial state anti-proton, where
both particle momenta are separated by $t$ and form a formal `quasi
two--particle' state $|p_1,-p_2,S; t\rangle$. These states are used to form the operator 
matrix elements. 
The correctness of this procedure within the
light--cone expansion relies, first, on the rapidity gap between
the outgoing proton and the remaining hadronic part with invariant
mass $M_X$ and, second, on the special property of matrix elements
of the contributing light-cone operators to contain no absorptive part.
Independently, one could argue that the corresponding matrix element is a
pure phenomenological quantity satisfying restrictions imposed by quantum field theory. 
The general structure of the scattering amplitude is completely determined by the 
off-cone structure of the twist-2 Compton operator (\ref{FTampl}), cf. \cite{gey1}.

The structure functions for the diffractive process
can thus be obtained by analyzing the absorptive part 
\begin{equation}
W_{\mu\nu} ={\sf Im} T_{\mu\nu}
\end{equation}
of the expectation value
\begin{equation}
\label{matr}
 T_{\mu\nu}\kln{x} = \Matel{p_1,-p_2, S; t}{\widehat T_{\mu\nu}(x)}{p_1,-p_2,S; t}\,,
\end{equation}
with the well-known operator $\widehat T_{\mu\nu}$ of (virtual) Compton scattering
 defined as
\begin{equation}
\label{int_input}
 \widehat T_{\mu\nu}(x) \equiv i R \, T\kle{J_\mu\kln{\frac{x}{2}} \,
                                    J_\nu\kln{-\frac{x}{2}} {\cal S}} \, .
\end{equation}
In \cite{BGR2006}, based on a general quantum field theoretic consideration of virtual
Compton scattering at twist 2\cite{Geyer:2004by,gey2,Geyer:1999uq}, we specified
the various terms which contribute to the general structure  of the hadronic tensor
$W_{\mu\nu}= {\sf Im} T_{\mu\nu}$ in case of deep--inelastic diffractive scattering.
As shown in \cite{Blumlein:1999sc,Blumlein:2001sb} the operator $\widehat T_{\mu\nu}$
in lowest order of the non--local light--cone expansion \cite{LCE} contains the 
vector or axial vector operators only.
The scattering amplitude is obtained by the Fourier transform of the operator
$\widehat T_{\kls{\mu\nu}}\kln{x}$ and forming the matrix element (\ref{matr}).
Here, we want to study its twist--2  contributions including  target mass
and finite momentum transfer corrections. This is obtained by {\sf harmonic extension} 
\cite{Geyer:1999uq,BT,Eilers:2004mp} of the twist--2 light-cone operators
to twist--2 off-cone operators
\cite{Geyer:2001qf}, leading to
\begin{align}
\label{FTampl}
    \widehat T_{\mu\nu}^{\twz}\kln{q}
         &=
   -\,e^2 \int \frac{\text{d}^4 \! x}{2\im\pi^2} \;
      \frac{\e^{\im qx}\,x_{\lambda} }{\kln{x^2- \im\epsilon}^2}
   \left\{ {S_{\mu\nu |}}^{ \alpha\lambda }\,
    O_\alpha^\twz(\kappa x ,- \kappa x )
     {+} {\epsilon_{\mu\nu}}^{\alpha\lambda}\,
    O_{5\,\alpha}^{\twz}(\kappa x ,-\kappa x )
    \right\} \,,
\end{align}
{
with 
\begin{eqnarray}
   O_\alpha^\twz(\kappa x ,- \kappa x )&=& 
     i[\overline \psi (\kappa x)\gamma_\alpha \psi(-\kappa x)
     -[\overline \psi (-\kappa x)\gamma_\alpha \psi(\kappa x)]^\twz, \nonumber \\
   O_{5\,\alpha}^\twz(\kappa x ,- \kappa x )&=& 
     [\overline \psi (\kappa x)\gamma_5 \gamma_\alpha \psi(-\kappa x)
     +[\overline \psi (-\kappa x)\gamma_5 \gamma_\alpha \psi(\kappa x)]^\twz, \nonumber 
\end{eqnarray}  
and }
$\kappa =1/2$. The matrix elements can be written in terms of vectors 
$\KI^a_{\mu,(5)}$ and 2--dimensional Fourier-integrals over partonic distributions
$f_{a(5)}(z_+,z_-,t)$ summing over $a$,
\begin{align}
\label{non1}
\langle p_1,-p_2;t|\,e^2\,O_{\mu}^{\twz}(\kappa  x,-\kappa   x)\,|p_1,-p_2;t\rangle
& = \KI^a_{\mu}(p_\pm) \!\int \!\frac{D{\mathbb Z}}{(2\pi)^4}\,
    e^{i\kappa  x( p_-z_- + p_+ z_+)}\,
    f_{\,a}(z_+ ,z_-,t)\,,\\
\label{non15}
\langle p_1,-p_2,S;t|\,e^2 \,O_{5\mu}^{\twz}(\kappa x, -\kappa  x)\,|p_1,-p_2,S;t\rangle
& =  \KI^a_{5\,\mu}(p_\pm,S) \!\int\! \frac{D{\mathbb Z}}{(2\pi)^4}\,
    e^{i\kappa  x( p_-z_- + p_+ z_+)}\,
    f_{5\,a}(z_+ ,z_-,t)\,, \nonumber  \\
\end{align}
which is defined as asymptotic expression on the light-cone at $x^2=0 $.

We choose as kinematic factors for the representation of the 
matrix element of the non-local operator  for the symmetric part (\ref{non1}) 
\begin{eqnarray}
\label{kinsym+}
\KI^{1 \, \mu} = p_+^\mu\,, \qquad \qquad
\KI^{2 \, \mu} = \pi_-^\mu \equiv p^\mu_- - {\eta}{p^\mu_+}\,,
\qquad
\end{eqnarray}
and for its antisymmetric part (\ref{non15})
\begin{eqnarray}
\label{kinasym+}
\KI^{1\, \mu}_{5} = S^\mu\,, \qquad
\KI^{2\, \mu}_{5} = p_+^\mu \,{(p_2 S)}/{M^2}\,, \qquad
\KI^{3\, \mu}_{5} = \pi_-^\mu\,{(p_2 S)}/{M^2}\,.
\end{eqnarray}
The normalization to $M^2$ in  (\ref{kinasym+}) 
is arbitrary and has to be arranged with the definition of the corresponding 
distribution 
functions $ f_{a,(5\,a)}(z_+,z_-)$, respectively. The corresponding 
Lorentz-invariant has to be formed out of the hadronic momenta, except the 
spin vector, since the polarization--symmetries are assumed to be linear in 
the spin.

The momentum fractions $z_{\pm}$ in (\ref{non1}, \ref{non15}) corresponding to the 
momenta $p_\pm$ are 
\begin{equation}
\mathbb P = (p_+, p_-)=(p_2+p_1,p_2-p_1), \,\quad
\mathbb Z = (z_+, z_-)=((z_2+z_1)/2,(z_2-z_1)/2)\,,
\end{equation}
with the measure $ D{\mathbb Z}$ 
\begin{eqnarray}
 D{\mathbb Z} = 2\, dz_+ dz_-\,
 \theta(1-z_++z_-)\,\theta(1+z_+-z_-)\,
 \theta(1-z_+-z_-)\,\theta(1+z_++z_-)\,.
 \end{eqnarray}

We refer to $f_{a(5)}(z_+,z_-,t)$ as {\sf diffractive generalized parton distribution 
functions, (dGPD),} in distinction to the GPDs emerging in deeply virtual Compton 
scattering~\cite{MRGDJ}. These amplitudes are directly connected to the total cross sections 
and polarization asymmetries, respectively. Both kinds of GPDs are expectation values of 
the same light--cone operator, however, between different states. 
Interesting limiting cases can be derived from them.
For the dGPDs these are the quasi collinear limit: $\pi_- 
\rightarrow 0, M^2 \rightarrow 0$, \cite{Blumlein:2001xf,Blumlein:2002fw}, and the limit 
of deep--inelastic scattering, see Appendix~B. Furthermore, for both types of GPDs the 
evolution equations are derived from the renormalization group equation for the {\sf same} 
light--cone operators. It is remarkable, that the evolution equations for the dGPDs are 
two--variable equations which reduce to the simple evolution equation for forward 
scattering in the quasi collinear limit, cf.~\cite{Blumlein:2001xf}. 

The (dimensionless) amplitudes $f_{(5)\,a}(z_+ ,z_-,t)$  depend  
on $t$ and $\eta$ explicitly.
In addition, there appears a $t$-- and $M^2$--dependence
of the amplitude (\ref{matr}) in momentum space, which finally, on the one hand,
results from the Fourier transform in (\ref{FTampl}) where the operator 
$ O_{(5)\,\alpha}^{\twz}(\kappa x ,-\kappa x )$ is off the light-cone, i.e.~with all 
trace subtractions. On the other hand, the dependence 
results from the kinematic 
pre--factors
$\KI^a_{(5)\,\mu}(p_\pm,S)$.\footnote{
In the following the explicit $t$--dependence of the distribution
functions is always understood and we drop this variable to lighten the 
notation.}

Concerning the independent kinematic factors one has 
two possibilities, which are  mathematically equivalent, depending on whether one chooses 
$p_-$ or 
$p_+$ as essential variable as we did in our previous papers \cite{BGR2006} and 
\cite{gey2}, respectively. The corresponding choices lead to different dGPDs.
\smallskip

\noindent
(1) In the first case, which we considered in \cite{BGR2006}, cf. also 
\cite{Blumlein:2001xf} and   \cite{Blumlein:2002fw},
 $p_-$ was chosen as essential variable,  by starting from 
the physical picture using the generalized optical theorem, 
and the parameterization~\footnote{
For later convenience the notation $(\vartheta, \zeta)$ of Ref.~\cite{BGR2006} has been 
changed into $(\hat\lambda,\hat\zeta)$.}
\begin{eqnarray}
 p_- z_- + p_+ z_+ &= \hat\lambda\,[p_- + \hat\zeta(p_+ - p_-/{\eta}) ] 
                   = \hat\lambda\,[p_- + \hat\zeta \hat\pi_-] 
                   \equiv \hat\lambda \,\hat{\cal P},
\end{eqnarray}
{ with}
\begin{eqnarray}
  \hat\lambda &=& z_- + z_+/{\eta}~, \nonumber\\
   z_+ &=& \hat\lambda \,\hat\zeta~,  \nonumber\\
   z_- &=& \hat\lambda\, (1 - \hat\zeta /{\eta})~.
\end{eqnarray}
\smallskip

\noindent
(2) A mathematically equivalent description is obtained starting from $p_+$ 
as the essential variable \cite{gey2}.  
In this approach we introduce the new variables $\lambda$ and $\zeta$
instead of $z_+$ and $z_-$, 
\begin{align}
\label{z+}
 p_- z_- + p_+ z_+ &= \lambda\,[\, p_+ + \zeta\,(p_- - {\eta} \,p_+)] = 
                   \lambda\, (p_+ + \zeta\, \pi_-) 
                   \equiv \lambda\, {\cal P} { = 2\,\Pi\,}, 
\end{align}
{with}
\begin{eqnarray}
   \lambda  &=&  z_+ + {\eta} z_-~, \nonumber\\
   z_- &=& \lambda \zeta~, \nonumber\\
   z_+ &=& \lambda (1 - \zeta {\eta})~.  
\end{eqnarray}
Here the variable $\lambda $  plays the role 
of a {\sf common} scale for $z_\pm $. 
Compared to Ref.~\cite{gey2} we list the essential kinematic variables using the 
above parameterization
\begin{eqnarray}
\label{cP2}
    {\cal P}(\eta , \zeta) &=& p_+ (1 - \eta\,\zeta) + p_- \zeta~, \\
    {\cal P}^2 &=& p_+^2 - 2\,\zeta\,\eta p_+^2  + \zeta^2 (p_-^2 +p_+^2 \eta^2),
\\
     q{\cal P} &=& \, qp_+ ,\nonumber\\
    {\cal P}^2/ ({\cal P}^{\mathrm T})^2 &=& x^2 ({\cal P}^2/Q^2)\,/\left[1 + x^2 ({\cal 
P}^2/Q^2)\right],
\end{eqnarray}
and
\begin{eqnarray}
\label{xicd}
 \xi_\pm
 =  \frac{2x}{1\pm \sqrt{1 + x^2 \mathcal{P}^2/Q^2}}\,,
 \quad x=\frac{Q^2}{qp_+} = \frac{Q^2}{2qp_1}(1-\eta) = x_{\rm BJ}(1-\eta) = - 2\, 
\beta\,\eta.
\end{eqnarray}
Obviously, $\xi_+ \equiv \xi $ is the appropriate generalization of the Nachtmann variable.
With these definitions the measure of the $\mathbb Z$--integration  is
\begin{align}
\label{varlz}
 D{\mathbb Z} 
 = 2|\lambda|\, d\lambda\, d\zeta \,
 &\theta\big(1-\lambda+(1+\eta)\lambda\,\zeta \big)\, 
 \theta\big(1+\lambda- (1+\eta)\lambda\,\zeta\big) \nonumber \\ 
 \times\,
 &\theta\big(1-\lambda-(1-\eta)\lambda\,\zeta \big)\,
 \theta\big(1+\lambda +(1-\eta)\lambda\,\zeta\big)\,~.
 \end{align}
In the present treatment we choose $p_+$ as the essential variable.

In Ref.~\cite{BGR2006}  deep--inelastic diffractive scattering has been worked out
within the first approach. The resulting expressions contain an internal
$\hat\zeta$--integral which is not well suited for the direct comparison of 
experimental data with 
the
diffractive GPDs. One way out is to introduce new `integrated distributions'.
Furthermore, we can perform a systematic $1/Q^2$ expansion which leads to an 
expansion in terms of ${\cal P}^2/Q^2$ directly. Since ${\cal P}^2$ is a 
polynomial of 
second order 
in the variable $\zeta $ we are led  to a $\zeta$-- expansion,
\begin{eqnarray}
 \hat{\cal P}^2 &=& t -2\,\hat\zeta\,t/\eta  + (4M^2 - t + t/\eta^2)\, 
 \hat{\zeta}^2|_{\hat\zeta \rightarrow 0} = t  \,,   \\
 {\cal P }^2 &=& \hat{\cal P}^2/\,\eta^2 = 
                  ( 4\,M^2 -t)(1 -2\,\eta\,\zeta) + \left[\,t + (4\,M^2 -t)\,\eta^2 \right] 
   \zeta^2|_{\zeta \rightarrow 0} = (4M^2 - t).
\label{psquared}
\end{eqnarray}
We prefer the second parameterization
which leads to expressions which contain as lowest approximation the mass corrections
known from deep--inelastic scattering, without requiring any further redefinition of the 
dGPDs.
We use the original expression for the
Compton scattering amplitude \cite{gey2} with the $\lambda$--parameterization and
apply the matrix elements (\ref{non1}, \ref{non15}).

\section{The Hadronic Tensor}
\renewcommand{\theequation}{\thesection.\arabic{equation}}
\setcounter{equation}{0}
\label{sec-unp}
In the following we discuss the symmetric and antisymmetric contributions
to the hadronic tensor, which correspond to the unpolarized and 
polarized case, separately.
\subsection{The Symmetric Part}

\vspace{1mm}
\noindent 
The symmetric part of the hadronic tensor for diffractive 
scattering, cf.~\cite{BGR2006,gey2} is given by
\begin{eqnarray}
W^\twz_{\{\mu\nu\}}\kln{q} &=& {\sf Im}~\frac{q^2}{2}\int D{\mathbb Z} \,
      \frac{{\cal A}_{\{\mu\nu\}}(q,{\cal P })}
      {\lambda \sqrt{(q{\cal P })^2- q^2 {\cal P }^2 }}
       \Bigg( \frac{1}{1 - \xi_+/\lambda + i \varepsilon}
         -  \frac{1}{1 - \xi_-/\lambda - i \varepsilon} \Bigg) \nonumber\\
&=&
- 2\pi\! \int\! d\zeta 
 \frac{q^2}{\sqrt{(q{\cal P })^2- q^2 {\cal P }^2 }}
     \Bigg\{
 \frac{q{\cal K}^a}{q{\cal P}}\; \bigg[
 g_{\mu\nu}^{\mathrm T} F_{a\, 1}(\xi,\zeta)
 -
 \frac{{\cal P}_\mu^{\mathrm T} {\cal P}_\nu^{\mathrm T}}
      {({\cal P}^{\mathrm T})^2}\,
 F_{a\, 2}(\xi,\zeta) \bigg]
 \nonumber \\
 && \hspace*{4.5cm}
 +\Big( \frac{q{\cal K}^a}{q{\cal P}}
     - \frac{{\cal P}{\cal K}^a}{{\cal P}^2} \Big)
 \bigg[
 g_{\mu\nu}^{\mathrm T} F_{a\, 3}(\xi,\zeta)
 - \frac{{\cal P}_\mu^{\mathrm T} {\cal P}_\nu^{\mathrm T}}
      {({\cal P}^{\mathrm T})^2}\,
 F_{a\, 4}(\xi,\zeta) \bigg]
\nonumber\\
 && \hspace*{4.5cm}
 -\bigg(\frac{{\cal K}_{\mu}^{a\,\mathrm T} {\cal P}_\nu^{\mathrm T}
 + {\cal P}_\mu^{\mathrm T} {\cal K}_{\nu}^{a\,\mathrm T}}
      {({\cal P}^{\mathrm T})^2}\,
 - 2\, \frac{q{\cal K}^a}{q{\cal P}}\,
      \frac{{\cal P}_\mu^{\mathrm T} {\cal P}_\nu^{\mathrm T}}
      {({\cal P}^{\mathrm T})^2} \bigg)
 F_{a\, 5}(\xi,\zeta)
\Bigg\}.\! \nonumber\\
\label{Ts_nonfim}
\end{eqnarray}
The relevant imaginary part belongs to the $\delta$--distribution $
\delta(1 - \xi_+/\lambda)$ in terms of variables $(\xi_+ \equiv \xi, \zeta)$, 
with the $\lambda$--integration, (\ref{varlz}), being carried out.
It implies the pole condition, cf.\cite{gey2} Eqs.~(6.6--6.10) and 
\cite{BGR2006},
\begin{align}
1 + \f12 \xi\,x\,{\cal P}^2/Q^2 = \sqrt{1+x^2{\cal P}^2/Q^2} = - (1-2\,x/\,\xi)~.
\label{pc}
\end{align} 
which we use below. The structure functions $F_{a i},i=1, \ldots, 5$ are given by
\begin{align}
\label{F1x}
F_{a\,1}(\xi,\,\zeta )
  &=
  \Phi^{(0)}_a(\xi,\zeta)
  + \frac{1}{2} \frac{x\,{\cal P}^2/Q^2}{\sqrt{1 + x^2{\cal P}^2/Q^2}}\;
  \Phi^{(1)}_a(\xi, \zeta)
  + \frac{1}{4} \frac{(x\,{\cal P}^2/Q^2)^2}{1 + x^2{\cal P }^2/Q^2}\;
  \Phi^{(2)}_a(\xi, \zeta) \,,
\\
F_{a\,2}(\xi,\, \zeta)
  &=
  \Phi^{(0)}_a(\xi,\zeta)
  + \frac{3}{2}
  \frac{x\,{\cal P}^2/Q^2}{\sqrt{1 + x^2{\cal P}^2/Q^2}}\;
  \Phi^{(1)}_a(\xi, \zeta)
  + \frac{3}{4}
  \frac{(x\,{\cal P}^2/Q^2)^2}{1 + x^2{\cal P}^2/Q^2}\;
  \Phi^{(2)}_a(\xi, \zeta) \,.
\label{F2x}
\end{align} 
\begin{eqnarray}
\label{F3x}
F_{a\,3}(\xi,\zeta) &=&
    -\frac{1}{2} \frac{\xi\,x\,{\cal P}^2/Q^2} {\sqrt{1+x^2{\cal P}^2/Q^2}}\, 
     \Phi^{(0)}_a(\xi,\zeta) \\
  && +\frac{1}{2\xi}
   		\bigg( \frac{\xi\,x\,{\cal P}^2/Q^2}{\sqrt{1+x^2{\cal P}^2/Q^2}}
    	- \frac{(\xi\,x\,{\cal P}^2/Q^2)^2}{1+x^2{\cal P}^2/Q^2} \bigg)  
     	\Phi_a^{(1)}(\xi,\zeta) 
\nonumber\\ 
 && -\frac{1}{\xi}
     \bigg( \frac{\xi\,x\,{\cal P}^2/Q^2} {\sqrt{1+x^2{\cal P}^2/Q^2}}
     -\frac{(\xi\,x{\cal P}^2/Q^2)^2} 
      {1+x^2{\cal P}^2/Q^2}
     +\frac{3}{8}\frac{(\xi\,x\,{\cal P}^2/Q^2)^3}
       {\sqrt{1+x^2{\cal P}^2/Q^2}^{\;3}}\bigg) 
      \int_\xi^1 \frac{dy}{y}\, \Phi_a^{(1)}(y,\zeta)\nonumber  \\
  && - \frac{1}{\xi}
    \bigg(\frac{(\xi\,x\,{\cal P}^2/Q^2)^2} {1+x^2{\cal P}^2/Q^2}
    -\frac{3}{4} \frac{(\xi\,x\,{\cal P}^2/Q^2)^3}
     {\sqrt{1+x^2{\cal P}^2/Q^2}^{\;3}}
     +\frac{3}{16} \frac{(\xi\,x\,{\cal P}^2/Q^2)^4}
     {[1+x^2{\cal P}^2/Q^2]^2}\bigg) 
      \int_\xi^1 \frac{dy}{y^2}\, \Phi_a^{(2)}(y,\zeta)\,, 
      \nonumber\\
\label{F4x}
F_{a\,4}(\xi,\zeta) &=&
    -\frac{1}{2}  \frac{\xi\,x{\cal P}^2/Q^2} {\sqrt{1+x^2{\cal P}^2/Q^2}} \, 
     \Phi^{(0)}_a(\xi,\zeta) \\
  && +\frac{1}{\xi}\bigg( \frac{5}{2} \frac{\xi\,x{\cal P}^2/Q^2}{\sqrt{1+x^2{\cal P}^2/Q^2}}
    - \frac{3}{2} \frac{(\xi\,x{\cal P}^2/Q^2)^2}{1+x^2{\cal P}^2/Q^2} \bigg)  
     \Phi_a^{(1)}(\xi,\zeta) \nonumber \\
  && -\frac{3}{\xi} \bigg( \frac{\xi\,x{\cal P}^2/Q^2} {\sqrt{1+x^2{\cal P}^2/Q^2}}
     -2\frac{(\xi\,x{\cal P}^2/Q^2)^2}{1+x^2{\cal P}^2/Q^2}
     +\frac{5}{8}\frac{(\xi\,x{\cal P}^2/Q^2)^3}
       {\sqrt{1+x^2{\cal P}^2/Q^2}^{\;3}}\bigg) 
      \int_\xi^1 \frac{dy}{y}\, \Phi_a^{(1)}(y,\zeta)\nonumber  \\
   &&  - \frac{3}{\xi}\bigg( 
     \frac{(\xi\,x{\cal P}^2/Q^2)^2} {1+x^2{\cal P}^2/Q^2}
     -\frac{5}{4} \frac{(\xi\,x{\cal P}^2/Q^2)^3}
       {\sqrt{1+x^2{\cal P}^2/Q^2}^{\;3}}
     +\frac{5}{16} \frac{(\xi\,x{\cal P}^2/Q^2)^4}
       {[1+x^2{\cal P}^2/Q^2]^2}\bigg) 
      \int_\xi^1 \frac{dy}{y^2}\, \Phi_a^{(2)}(y,\zeta)\, ,
      \nonumber\\
F_{a\, 5}(\xi,\zeta) &=&  \frac{1}{\xi}\bigg[\Phi_a^{(1)}(\xi,\zeta) 
  + \frac{3}{2}\frac{\xi\,x{\cal P}^2/Q^2}{\sqrt{1 + x{\cal P}^2/Q^2}} 
        \int_\xi^1 \frac{dy}{y}\,\Phi_a^{(1)}(y,\zeta) 
        \label{F5x}\nonumber\\
  && +  \frac{3}{4}\frac{(\xi\,x{\cal P}^2/Q^2)^2}{1 + x{\cal P}^2/Q^2} 
        \int_\xi^1 \frac{dy}{y^2}\,\Phi_a^{(2)}(y,\zeta)\bigg] \,.   
\end{eqnarray}
Whereas $F_{a\, 1(2)}(\xi,\zeta)$  are direct generalizations of the well-known 
deep--inelastic structure functions. $F_{a\, k}(\xi,\zeta)|_{k=3,4,5}$ are
new structure functions, which vanish in the forward limit, cf.~Appendix B.
The typical square roots ${\sqrt{1 + x^2{\cal P}^2/Q^2}} $ for the mass corrections 
depend on the generalized momentum ${\cal P} = {\cal P}(\zeta).$
After substituting 
$\lambda \rightarrow \xi $ 
in (\ref{F1x}--\ref{F5x}), 
we introduce the following iterated representations for the basic 
dGPDs $f_a (\lambda,\zeta)$, cf.~(\ref{non1}):
\begin{align}
\label{phia0}
\Phi_a^{(0)}(\xi,\zeta) &\equiv f_a (\xi,\zeta)\,,
\\
\label{phia1}
\Phi_a^{(1)}(\xi,\zeta)
&\equiv
      \int_\xi^{1} dy_1 \;f_a (y_1,\zeta)
 = \xi \int_0^1 \frac{d\tau}{\tau^2}\;
   f_a \Big(\frac{\xi}{\tau},\zeta\Big)\,,
\\
\label{phia2}
\Phi_a^{(2)}(\xi,\zeta)
&\equiv
\int_\xi^{1} dy_2
\int_{y_2}^{1} dy_1 \, f_a (y_1,\zeta)
= \xi^2 \int_0^1 \frac{d\tau_1}{\tau_1^3}
  \int_0^1 \frac{d\tau_2}{\tau_2^2}\;
   f_a \Big(\frac{\xi}{\tau_1\tau_2},\zeta\Big)\,,
\\
\label{eqphiaa}
\Phi_a^{(i)}(\xi,\zeta)
&\equiv
\int_\xi^{1} dy\; \Phi_a^{(i-1)}(y,\zeta)\, , \qquad \mbox{for}\quad i \geq 1 \,, \\
\label{eqphiaa1}
\int_\xi^1 \frac{dy}{y}\,\Phi_a^{(1)}(y,\zeta)  
 &\equiv \int_\xi^1 \frac{dy_1}{y_1}\int_{y_1}^1\!\! dy\, \Phi_a^{(0)}(y,\zeta) 
 = \xi\!\int_0^1 \frac{d\tau_1}{\tau_1^2}\int_0^1 \frac{d\tau_2}{\tau_2^2}\;
   f_a \Big(\frac{\xi}{\tau_1\tau_2},\zeta\Big)\,,
\\
\label{eqphiaa2}
\int_\xi^1 \frac{dy}{y^2}\,\Phi_a^{(2)}(y,\zeta)  
 &\equiv \int_\xi^1 \frac{dy_1}{y_1^2}\int_{y_1}^1
 \!\! dy\, \Phi_a^{(1)}(y,\zeta)  
   =    \xi\! \int_0^1 \frac{d\tau_1}{\tau_1^3}
  \int_0^1 \frac{d\tau_2}{\tau_2^2}\int_0^1 \frac{d\tau_3}{\tau_3^2}    \;
   f_a \Big(\frac{\xi}{\tau_1\tau_2\tau_3 },\zeta\Big).
\end{align}

\vspace{2mm}\noindent
Let us now investigate the effect of target masses and finite terms in $t$ in more 
detail. It turns out that both the $M^2$-- and $t$--contributions in the diffractive 
structure 
functions emerge due to the parameter $\rho$
\begin{eqnarray}
\label{rho}
\rho =  \epsilon \, x^2 \frac{p_+^2}{Q^2} \frac{1}{1+x^2 p_+^2/Q^2}~,
\end{eqnarray}
with  $\epsilon$ given by ${\cal P}^2 = p_+^2(1+\epsilon)$,
\begin{eqnarray}
\label{epsi}
\epsilon = \frac{1}{p_+^2} \left[2 \zeta p_+ \pi_- + \zeta^2 \pi_-^2 \right] = - 2 \eta \zeta
           +\left(\eta^2 + \frac{t}{p_+^2}\right) \zeta^2~.
\end{eqnarray}
Since
\begin{eqnarray}
-\eta \simeq x_\PP~ \ll 1~,
\end{eqnarray}
$\rho$ effectively takes values $\rho \lesssim 10^{-3}$ for $x_\PP \lesssim 10^{-2}, |t| 
\approx 
(0.1 ... 1 ) M^2, Q^2 \approx (1 ... 5 ) M^2.$ The range of $\zeta$ is 
determined both 
by the support condition (\ref{varlz}) and the condition $ {\cal P}^2 = 
p_+^2(1+ \epsilon) > 0 $ in the diffractive case.

To prepare the expansion in $\rho$ we rewrite the hadronic tensor as 
\begin{align}
W^\twz_{\{\mu\nu\}}\kln{q}
 =
  2 \pi \!&\int\! d\zeta\;
   \frac{q{\cal K}^a}{q\cal P}\,
  \bigg[
 - g_{\mu\nu}^{\mathrm T}\, {W}^{\rm diff}_{a\,1} 
     \Big(x,\frac{{\cal P}^2}{Q^2};\zeta\Big) 
 + \frac{{\cal P}_\mu^{\mathrm T} {\cal P}_\nu^{\mathrm T}}{M^2}
 \,{W}^{\rm diff}_{a\,2}\Big(x,\frac{{\cal P}^2}{Q^2};\zeta\Big) \bigg]\,,
 \nonumber
 \\ 
& +
 \Bigg\{\!
 \bigg( \frac{q{\cal K}^a}{q{\cal P}}
     - \frac{{\cal P}{\cal K}^a}{{\cal P}^2} \bigg)\frac{{\cal P}^2}{Q^2}
 \bigg[
 - g_{\mu\nu}^{\mathrm T}{W}^{\rm diff}_{a\,3} \Big(x,\frac{{\cal P}^2}{Q^2};\zeta\Big) 
 + \frac{{\cal P}_\mu^{\mathrm T}{\cal P}_\nu^{\mathrm T}}{M^2}
  {W}^{\rm diff}_{a\,4} \Big(x,\frac{{\cal P}^2}{Q^2};\zeta\Big)
  \bigg]
  \nonumber\\
&  +
 \bigg({\cal P}_\mu^{\mathrm T}{\cal K }_\nu^{a\,\mathrm T}
       + {\cal P}_\nu^{\mathrm T}{\cal K }_\mu^{a\,\mathrm T}
       - 2\,
 {\cal P}_\mu^{\mathrm T}{\cal P}_\nu^{\mathrm T} \frac{q {\cal K}^a}{q{\cal P}}
  \bigg)
 \frac{1}{M^2}
   {W}^{\rm diff }_{a\,5}\Big(x,\frac{{\cal P}^2}{Q^2};\zeta\Big) 
  \Bigg\}\,~. 
\label{TSext}
\end{align}
The integral over $\zeta$ cannot be  performed easily. Here, the un-integrated 
structure functions ${W}^{\rm diff}_{a\,k }(x,{{\cal P}^2(\zeta)}/{Q^2};\zeta)$ are given by
\begin{align}
     {W}^{\rm diff}_{a\,1 }\Big(x,\frac{{\cal P}^2(\zeta)}{Q^2};\zeta\Big)
  & \equiv
   - \frac{x}{\sqrt{1+x^2\mathcal{P}^2/Q^2}}
    \; F_{a\,1}(\xi,\zeta)~, 
    & & 
\label{WF1}
    \\
     {W}^{\rm diff}_{a\,3 }\Big(x,\frac{{\cal P}^2(\zeta)}{Q^2};\zeta\Big)
  & \equiv
   - \frac{x}{\sqrt{1+x^2\mathcal{P}^2/Q^2}} \frac{Q^2}{{\cal P}^2}
    \; F_{a\,3}(\xi,\zeta)~, 
    & &
\label{WF3}
    \\
     {W}^{\rm diff}_{a\,k }\Big(x,\frac{{\cal P}^2(\zeta)}{Q^2};\zeta\Big) 
  & \equiv
 - \frac{M^2}{Q^2}\,
   \bigg(\frac{x}{\sqrt{1+x^2\mathcal{P}^2/Q^2}} \bigg)^{\!\!3}
    F_{a\,k}(\xi,\zeta) 
    &  &\mbox{for~} k= 2,5\;,
\label{WFj}
     \\ 
     {W}^{\rm diff}_{a\,4 }\Big(x,\frac{{\cal P}^2(\zeta)}{Q^2};\zeta\Big) 
  & \equiv
 - \frac{M^2}{{\cal P }^2}\,
   \bigg(\frac{x}{\sqrt{1+x^2\mathcal{P}^2/Q^2}} \bigg)^{\!\!3}
    F_{a\,4}(\xi,\zeta)~. 
    &  & 
\label{WF4}
\end{align}
As noted in \cite{BGR2006}
a generalized Callan--Gross \cite{CG} relation between ${W}_{a\,1}^{\rm diff}$ and
${W}_{a\,2}^{\rm diff}$, which holds for diffractive scattering in the limit
$M^2,~t~\rightarrow~0$, \cite{Blumlein:2001xf}, is broken as in the case of
deep--inelastic scattering \cite{Georgi:1976ve}.
Correspondingly, the distribution 
functions
${W}_{a\,(1,2)}^{\rm diff}$ are related to ${W}_{a\, \rm L}^{\rm diff}$, the diffractive
analogue of the longitudinal structure function of deep--inelastic scattering, by
\begin{align}
{W}^{\rm diff }_{a\,\rm L}(x,{\cal P}^2/Q^2;\zeta)
&=
- \,{W}^{\rm diff }_{a\,1}(x,{\cal P}^2/Q^2;\zeta)
+  \left(1+\frac{x^2{\cal P}^2}{Q^2}\right)\frac{qp_+}{x\, M^2}\,
 {W}^{\rm diff }_{a\,2}(x,{\cal P}^2/Q^2;\zeta)\,.
 \label{w1}
\intertext{
To see this in detail, we insert (\ref{F1x}), (\ref{F2x})  and  (\ref{WFj}), so that}
{W}^{\rm diff }_{a\,\rm L}(x,{\cal P}^2/Q^2;\zeta)
&= \frac{x}{\sqrt{1 + x^2{\cal P}^2/Q^2 }}
     \big( F_{a\,1}(\xi,\zeta)  -  F_{a\,2}(\xi,\zeta) \big) \approx  
{\cal O}\left(\frac{x^2{\cal P}^2}{Q^2}\right). 
\label{CGk}
\intertext{
The last relation follows by direct inspection of $ F_{a\,i}$ and is explicit in}
  {W}^{\rm diff }_{a\,\rm L}(x,{\cal P}^2/Q^2;\zeta)& =
      -\frac{x^2 {\cal P}^2}{2Q^2}\frac{\partial}{\partial {x}}
       \Big(\frac{x}{\xi \sqrt{1 + x^2{\cal P}^2/Q^2 }}
        \Phi_a^{(2)}(\xi,\zeta)\Big)
\end{align}
derived in \cite{gey2}, cf. also \cite{Georgi:1976ve,Blumlein:1998nv} for the case of 
forward scattering.

Most of the above quantities depend on ${\cal P}^2$ (\ref{cP2})  which we write now as
\begin{eqnarray}
\label{psquared1}
{\cal P}^2 &=& p_+^2 + 2\zeta p_+ \pi_- +\zeta^2 \pi_-^2 =  p_+^2 (1+ \epsilon)\,~.  
\end{eqnarray}
Let us simplify the contraction of the kinematic coefficients in (\ref{TSext}).
For ${\cal K}_1 = p_+ $, observing $q{\cal P} = qp_+$ and Eq.~(\ref{psquared1}) 
for ${\cal P}^2$, we obtain
\begin{align}
\label{kin11}
 \frac{q{\cal K}_1}{q{\cal P}} &= 1\,, \\
\label{kin21}
 \Big(\frac{q{\cal K}_1}{q{\cal P}} - \frac{{\cal P}{\cal K}_1}{{\cal P}^2}\Big)
  \frac{{\cal P}^2}{Q^2}
 & = \frac{p_+^2}{Q^2}\bigg(-\eta \zeta  + \zeta^2\left(\eta^2 + \frac{t}{p_+^2}\right) \bigg)
   \\
\label{kin31}
{\cal K}_{1\mu}^{\mathrm T} {\cal P}_\nu^{\mathrm T}
 + {\cal P}_\mu^{\mathrm T} {\cal K}_{1\nu}^{\mathrm T}
 -2\frac{q{\cal K}_1}{q{\cal P}}
 {\cal P}_\mu^{\mathrm T} {\cal P}_\nu^{\mathrm T}
 &=
 -\,\zeta\left(p_{+\mu}^{\mathrm T}\pi_{-\nu} + p_{+\nu}^{\mathrm T}\pi_{-\mu}\right)
 - 2\,\zeta^2 \pi_{-\mu}\pi_{-\nu}\,,
 \\
\intertext{and for ${\cal K}_2 = \pi_- $, due to the transversality of $\pi_-$, 
one finds}
\label{kin12}
 \frac{q{\cal K}_2}{q{\cal P}} &= 0\,, \\
\label{kin22}
\Big( \frac{q{\cal K}_2}{q{\cal P}} - \frac{{\cal P}{\cal K}_2}{{\cal P}^2}\Big)
    \frac{{\cal P}^2}{Q^2}
& = \frac{p_+^2}{Q^2}\bigg( \eta -\zeta\left(\eta^2 + 
\frac{t}{p_+^2}\right)\bigg) , \\
\label{kin32}
{\cal K}_{2\mu}^{\mathrm T} {\cal P}_\nu^{\mathrm T}
 + {\cal P}_\mu^{\mathrm T} {\cal K}_{2\nu}^{\mathrm T}
 -2\frac{q{\cal K}_2}{q{\cal P}}
 {\cal P}_\mu^{\mathrm T} {\cal P}_\nu^{\mathrm T}
 &=
 \left(p_{+\mu}^{\mathrm T}\pi_{-\nu} + p_{+\nu}^{\mathrm T}\pi_{-\mu}\right)
 + 2\,\zeta \,\pi_{-\mu}\pi_{-\nu}\,.
 \end{align}
It is remarkable that only for ${\cal K}_1 = p_+ $  the first invariant
${q{\cal K}_1}/{q{\cal P}}$  contributes to the zeroth power in $\zeta$, 
whereas the other ones start at most with the {first} power.  
The contributions of  invariants belonging to kinematic coefficients
containing  $\pi_-$  are less important because this variable is transverse
to $q$ with $\pi_- q =0$. The corresponding invariants
\begin{equation}
\pi_-^2 = t +\eta^2\, p_+^2,\; \pi_-p_+ = -\eta \,p_+^2,\; \pi_-p_- = t~,
\end{equation}
are small compared to $Q^2$.

Having now expressed the $\zeta$--dependence in all kinematic factors explicitly,
we may perform the $\zeta$--integral introducing $n$th moments~: 
\begin{eqnarray}
\label{mg}
 {W}^{(n)\,\rm diff}_{a\,k}(x,\eta,t,p_+^2/Q^2)= \int\! d\zeta\, \zeta^n 
           \,{W}^{\,\rm diff}_{a\,k}(x,{\cal P}^2/Q^2;\zeta)\,~.
\end{eqnarray}
The hadronic tensor reads
\begin{align}
\frac{1}{2\pi}{\sf Im}\,&T^\twz_{\{\mu\nu\}}\kln{q}
  =\nonumber  \\ 
 &  - g_{\mu\nu}^{\mathrm T}\,
    \left\{\, {W}^{(0)\,\rm diff}_{1\,1} + \frac{p_+^2}{Q^2}\left[
             \eta \, ({W}^{(0)\,\rm diff}_{2\,3}
           - {W}^{(1)\rm diff}_{1\,3}) +\left(\eta^2 
+\frac{t}{p_+^2}\right)\,({W}^{(2)\rm diff}_{1\,3}
           - {W}^{(1)\rm diff}_{2\,3}) \right]
         \right\}\nonumber \\
 & + \frac{p_{+\,\mu}^{\mathrm T} p_{+\,\nu}^{\mathrm T}}{M^2} \left\{
      {W}^{(0)\,\rm diff}_{1\,2} +
      \frac{p_+^2}{Q^2} \left[\eta \, \left({W}^{(0)\,\rm diff}_{2\,4}
      -{ {W}^{(1)\,\rm diff}_{1\,4}}\right) +
      \left(\eta^2 + \frac{t}{p_+^2}\right)\,( {W}^{(2)\,\rm diff}_{1\,4} -
      {W}^{(1)\,\rm diff}_{2\,4})
         \right]
          \right\} \nonumber 
\end{align}\begin{align}
 &  +\frac{p_{+\,\mu}^{\mathrm T} \pi_{-\,\nu}+ p_{+\,\nu}^{\mathrm T} \pi_{-\,\mu}}
     {M^2}\left\{{W}^{(1)\,\rm diff}_{1\,2} + 
        \frac{p_+^2}{Q^2} \Bigl[\eta ({W}^{(1)\,\rm diff}_{2\,4}
         - {W}^{(2)\,\rm diff}_{1\,4})  \right.
\nonumber  \\
 & \qquad \qquad \qquad \qquad \left.
   +\left(\frac{t}{p_+^2} +\eta^2\right)({W}^{(3)\,\rm diff}_{1\,4}
-{W}^{(2)\,\rm diff}_{2\,4})\Bigr]
     +{{W}^{(0)\,\rm diff}_{2\,5}} -  {W}^{(1)\,\rm diff}_{1\,5} 
\right\} 
\nonumber \\
 &  +  \frac{\pi_{-\,\mu} \pi_{+\,\nu}}{M^2}\,
           \left\{{W}^{(2)\,\rm diff}_{1\,2} + 2{W}^{(1)\,\rm diff}_{2\,5}
            -2 {W}^{(2)\,\rm diff}_{1\,5} \right.\nonumber \\ 
& \quad \qquad  \left. 
  + \frac{p_+^2}{Q^2}
\left[
     \eta \, \left(
             {W}^{(2)\,\rm diff}_{2\,4}- {W}^{(3)\,\rm diff}_{1\,4}\right)  
           + \left(\eta^2 + \frac{t}{p_+^2} \right)
             \left( {W}^{(4)\,\rm diff}_{1\,4}- {W}^{(3)\,\rm diff}_{2\,4}  
             \right) 
\right]
\right\}~.    
\label{WPhi0} 
\end{align}
Here the momentum fraction argument of the structure functions 
 $W_{ak}^{\rm diff}$  is the original Nachtmann variable (\ref{xicd}), 
whereas for the functions ${W}^{(n)\,\rm diff}_{a\,k}(x,\eta,t,p_+^2/Q^2)$ it is $x$.
These structure functions are in principle accessible experimentally, 
varying the  
external kinematic parameters $x_{\rm BJ}, Q^2, t$ and $x_\PP, (\eta = 
\eta(x_\PP))$. 

Up to this point no approximations have been made. We would now like to 
discuss the above structure.
Note that  disregarding of $\pi_-$ as transversal degree of freedom corresponds
to the limit  $ \epsilon \rightarrow 0 $, (\ref{psquared1}). However, $\epsilon$, 
(\ref{epsi}), is not necessarily a small quantity. 
The Taylor expansion in $\epsilon$ would retain the DIS-like  target mass corrections
and lead to the physically relevant  power series in $p_+^2/Q^2$ of the denominators. 
Because of the smallness of $\rho $ (\ref{rho}) and also 
$x^2 p_+^2/Q^2 $ we use the latter as expansion parameter.
Thereby the Nachtmann variable is substituted by $x$ in lowest order,
whereas by setting $\pi_-=0$  we would retain an approximate Nachtmann variable, 
\begin{eqnarray}
\label{zeta0}
\xi_0= 2x/(1+\sqrt{1 +x^2p_+^2/Q^2}). 
\end{eqnarray}
For simplicity we proceed as follows:
\begin{itemize}
\item effective expansion w.r.t. the parameter $p_+^2/Q^2$,
 \begin{eqnarray}
 \label{epsilon2}
 \bigg(1+ x^2 \,\frac{{\cal P}^2}{Q^2}\bigg)^{-n} =
 \bigg(1+ x^2 \,\frac{p_+^2(1+\epsilon)}{Q^2}\bigg)^{-n} =
\left(1- n x^2\,\frac{ p_+^2}{Q^2}(1+\epsilon) + ...  \right),
\end{eqnarray}
\item expansion of the Nachtmann variable (\ref{xicd}),
\begin{eqnarray}
\label{xi_00}
  \xi- x = -\frac{1}{4}x \frac{x^2p_+^2}{Q^2}(1+\epsilon) + ...\, ,
\end{eqnarray}
\item use of $x$ instead of the Nachtmann variable $\xi$. 
\item For the treatment of the denominators we shift the integration variable
      $ \lambda = \lambda' + \xi - x$, 
 \begin{eqnarray}
  \frac{1}{\lambda -\xi + i\varepsilon \lambda} =\frac{1}{\lambda' -x + i\varepsilon \lambda} .
 \end{eqnarray}
\end{itemize}
Through this procedure we avoid the expansion of the denominator
 in favor of an expansion of the  dGPDs. In principle problems could arise because
 of possible differences in $ \varepsilon(\lambda - \lambda')$. 
Therefore we have to expand the basic  dGPD
\begin{eqnarray}
\Phi^{(i)}_a(\lambda, \zeta)& =&
\Phi^{(i)}_a(\lambda' +\xi-x , \zeta) 
   = \Phi^{(i)}_a(\lambda' , \zeta)
	+ \partial_{\lambda'}  \Phi^{(i)}_a(\lambda', \zeta) 
          (\xi - x) + \ldots \nonumber \\ 
 &=&\Phi^{(i)}_a(\lambda' , \zeta)   
    -\frac{1}{4}x \frac{x^2 p_+^2}{Q^2}(1+\epsilon) \partial_{\lambda'}  
     \Phi^{(i)}_a(\lambda', \zeta) + \ldots\quad. 
\nonumber \end{eqnarray}
As a test we can study the limit of 
deep--inelastic scattering, whereby we reproduce the standard result. For
diffractive DIS it is sufficient to consider the lowest approximation which 
extends our results \cite{Blumlein:2001xf,Blumlein:2002fw}. 
In the following we define moments of the dGPDs by
\begin{eqnarray}
\label{mph}
   \Phi_a^{(i\,{n})}(x) = \int d\zeta  \zeta^n \Phi_a^{(i)}(x,\zeta).
\end{eqnarray}
This corresponds to a change from a GPD to a parton density. 

Now we apply our approximation procedure directly to Eq.~(\ref{TSext}) using 
(\ref{WF1}--\ref{WF4}) and (\ref{F1x}--\ref{F5x}). We write the result separately for the
invariants ${\cal K}^1 = p_+$,
\begin{align}
\frac{1}{2\pi}{\sf Im}\, T^\twz_{\{\mu\nu\}}\kln{q}|_1
 & =    g_{\mu\nu}^{\mathrm T}\,\Bigg[  
         x \Phi_1^{(0\,0)}(x) + \frac{x^2 p_+^2}{Q^2}
         \Big(t_{1\,1}^0 + t_{1\,3}^0 - \tilde t_{1\,3}^0 + \eta \tilde t_{1\,3}^1
         \Big) \Bigg] 
 \nonumber \\
 & - \frac{p_{+\,\mu}^{\mathrm T} p_{+\,\nu}^{\mathrm T}}{Q^2} 
          \Bigg[x^3 \Phi_1^{(0\,0)} + \frac{x^2 p_+^2}{Q^2}   
          x^2\Big(t_{1\,2}^0 + t_{1\,4}^0 - \tilde t_{1\,4}^0 + \eta \tilde t_{1\,4}^{\,1}
           \Big) \Bigg]  
 \nonumber \\ 
 &  -\frac{p_{+\,\mu}^{\mathrm T} \pi_{-\,\nu}+ p_{+\,\nu}^{\mathrm T} \pi_{-\,\mu}}
     {Q^2}\Bigg[ x^3 \Phi_1^{(0\,1)} - x^2 \Phi_1^{(1\,1)}    
\nonumber\\ & \hspace*{4cm} 
+ \frac{x^2 p_+^2}{Q^2}x^2
     \Big(t_{1\,2}^1 + t_{1\,4}^1 - \tilde t_{1\,4}^{\,1} - t_{1\,5}^1   + 
      \eta \tilde t_{1\,4}^{\,2}
           \Big) \Bigg] 
 \nonumber \\ 
 &  -  \frac{\pi_{-\,\mu} \pi_{-\,\nu}}{Q^2}\,
           \bigg[ x^3 \Phi_1^{(0\,2)} - 2x^2 \Phi_1^{(1\,2)}+ \frac{x^2 p_+^2}{Q^2}x^2
     \Big(t_{1\,2}^2 + t_{1\,4}^2 - \tilde t_{1\,4}^{\,2} -2 t_{1\,5}^2   + 
      \eta \tilde t_{1\,4}^{\,3}
           \Big) \Bigg]~, 
\label{Wa1}
\intertext{and for ${\cal K}^2 = \pi_- $,  }
\frac{1}{2\pi}{\sf Im}\, T^\twz_{\{\mu\nu\}}\kln{q}|_2
 & =   g_{\mu\nu}^{\mathrm T}\,  
         \frac{x^2 p_+^2}{Q^2}
         \Big[\eta \tilde t_{2\,3}^{\,0} - \left(\eta^2 
+\frac{t}{p_+^2}\right)\tilde t_{2\,3}^{\,1}
          \Bigg]
 \nonumber \\
 &           - \frac{p_{+\,\mu}^{\mathrm T} p_{+\,\nu}^{\mathrm T}}{Q^2} 
          \frac{x^2 p_+^2}{Q^2}   
          x^2\Big[\eta \tilde t_{2\,4}^{\,0} 
          -\left(\eta^2 +\frac{t}{p_+^2}\right)\tilde t_{2\,4}^{\,1}  
           \Big) \Bigg]  
 \nonumber \\ 
 &  -\frac{p_{+\,\mu}^{\mathrm T} \pi_{-\,\nu}+ p_{+\,\nu}^{\mathrm T} \pi_{-\,\mu}}
     {Q^2}\Bigg[ x^2 \Phi_2^{(1\,0)}  + \frac{x^2 p_+^2}{Q^2}x^2
     \Big(t_{2\,5}^0 +{\eta t_{2\,4}^1 -\left(\eta^2 +\frac{t}{p_+^2}\right)
      \tilde t_{2\,4}^{\,2}} 
           \Big) \Bigg] 
 \nonumber \\ 
 &    -  \frac{\pi_{-\,\mu} \pi_{-\,\nu}}{Q^2}\,
           \bigg[2 x^2 \Phi_2^{(1\,1)} + \frac{x^2 p_+^2}{Q^2}x^2
     \Big( 2t_{2\,5}^1 + \eta t_{2\,4}^2 -\left(\eta^2 +\frac{t}{p_+^2}\right)   
         \tilde t_{2\,4}^{\,3}            \Big) \Bigg] .
\label{Wa2}
\end{align}
Here $t_{ai}^n$ and $\tilde t_{ai}^n$ are given by
\begin{align}
t_{a1}^n & = \int d\zeta (1+\epsilon(\zeta)) \zeta^n    
          \left(-\f12 x \Phi^{(0)}_a(x, \zeta) + \f12 \Phi^{(1)}_a(x, \zeta)
          -\frac{1}{4}x^2 \partial_{x}\Phi^{(0)}_a(x, \zeta)\right),
\nonumber \\
t_{a2}^n & = \int d\zeta (1+\epsilon(\zeta)) \zeta^n    
          \left(-\frac{3}{2} x\Phi^{(0)}_a(x, \zeta) + 
          \frac{3}{2}  \Phi^{(1)}_a(\xi_{00}, \zeta)
          -\frac{1}{4} x^2 \partial_{x}\Phi^{(0}_a(x, \zeta)\right),
\\
t_{a3}^n & = \int d\zeta (1+\epsilon(\zeta)) \zeta^n    
          \left(-\f12 x \Phi^{(0)}_a(x, \zeta) + 
          \frac{1}{2} \Phi^{(1)}_a(\xi_{00}, \zeta)
          -\int_x^1\frac{dy}{y}\Phi^{(1)}_a(y, \zeta)\right),
\nonumber 
\end{align}\begin{align}
\nonumber \\
t_{a4}^n & = \int d\zeta (1+\epsilon(\zeta)) \zeta^n    
          \left(-\f12 x\Phi^{(0)}_a(x, \zeta) + 
          \frac{5}{2} \Phi^{(1)}_a(x, \zeta)
          -3\int_x^1\frac{dy}{y}\Phi^{(1)}_a(y, \zeta)\right),
\nonumber \\
t_{a5}^n & = \int d\zeta (1+\epsilon(\zeta)) \zeta^n    
          \left(-\frac{5}{4} \Phi^{(0)}_a(x, \zeta) + 
          \frac{3}{2} \int_x^1\frac{dy}{y}\Phi^{(1)}_a(y, \zeta)
          -\frac{1}{4} x \partial_{x}\Phi^{(1)}_a(x, \zeta)\right),
\end{align}
and
\begin{align}
\tilde t_{a1}^{\,n} & = \int d\zeta  \zeta^n    
          \left(-\f12 x \Phi^{(0)}_a(x, \zeta) + \f12 \Phi^{(1)}_a(x, \zeta)
          -\frac{1}{4}x^2 \partial_{x}\Phi^{(0)}_a(x, \zeta)\right)~.
\end{align}
Similar
for all other terms $\tilde t_{ak}^n$, the factor $(1+\epsilon(\zeta)) $
           is absent compared to  $t_{ak}^n$.  

It is remarkable that each kinematic coefficient ${\cal K }^a$ contributes to all possible
kinematic  structures. Because of the transversal behaviour of $\pi_-$
we expect that the last two structures 
${p_{+\,\mu}^{\mathrm T} \pi_{-\,\nu}+ p_{+\,\nu}^{\mathrm T} \pi_{-\,\mu}}$ and
${\pi_{-\,\mu} \pi_{-\,\nu}}$ as well as the complete contributions of the second
invariant (\ref{Wa2})   are less important in comparison with the structures
$ g_{\mu\nu}^{\mathrm T}\,$ and  ${p_{+\,\mu}^{\mathrm T} p_{+\,\nu}^{\mathrm T}}$
and the first invariant in (\ref{Wa1}). Moreover the leading contributions to the first two
structures in (\ref{Wa2}) contain the small coefficient {$\eta$}. 
In Eqs.~(\ref{Wa1}, \ref{Wa2}) the
contributions $\propto  M^2, t$ emerge as	
\begin{eqnarray}
\frac{x^2 p_+^2}{Q^2} &=& \frac{x^2\left(4 M^2 - t\right)}{Q^2}~,\\
\frac{x^2 p_-^2}{Q^2} &=& \frac{x^2 t}{Q^2}~, 
\end{eqnarray}
respectively. Noting that $|\eta| \simeq x_\PP$, and $x_\PP \sim {\cal O}(x_{\rm BJ})$ for 
diffractive 
scattering the target mass and finite $t$ corrections are suppressed by 
${\cal O}(x^2 M^2/Q^2)$, with $ x \lesssim 10^{-2}$. In the meson-exchange case, 
$x$--values of around $x \simeq 0.3$ 
may be reached and ${\cal O}(10 \% \times (M^2/Q^2))$  
effects may be obtained.

Let us consider the complete zeroth order term 
\begin{align}
\frac{1}{2\pi}{\sf Im}\, T^\twz_{\{\mu\nu\}}\kln{q}|_0
  = & g_{\mu\nu}^{\mathrm T}\, \,x \Phi_1^{(0\,0)}(x)
       - \frac{p_{+\,\mu}^{\mathrm T} p_{+\,\nu}^{\mathrm T}}{Q^2} 
               x^3\,   \Phi_1^{(0\,0)} 
        \nonumber \\
 &  -\frac{p_{+\,\mu}^{\mathrm T} \pi_{-\,\nu}+ p_{+\,\nu}^{\mathrm T} \pi_{-\,\mu}}
     {Q^2} x^2\bigg[  x\,  \Phi_1^{(0\,1)} +
            \Phi_2^{(1\,0)}- \Phi_1^{(1\,1)} \bigg] 
 \nonumber \\
 &  -  \frac{\pi_{-\,\mu} \pi_{-\,\nu}}{Q^2}\,x^2  
           \bigg[ x\,\Phi_1^{(0\,2)} - 2\Phi_1^{(1\,2)} + 2 \Phi_2^{(1\,1)}
            \bigg]~.
\label{T_01}
\end{align}
Also here we can see that the contributions to the first two  kinematic structures 
result from the distribution functions  $\Phi_1^{(0\,0)}(x) $   of the first kinematic 
structure only. This reproduces our result \cite{Blumlein:2001xf} obtained for 
vanishing $t$, target mass,
and negligible transversal momenta $\pi_-$,
\begin{align}
\frac{1}{2\pi}{\sf Im}\, T^\twz_{\{\mu\nu\}}\kln{q}|_0
  =  g_{\mu\nu}^{\mathrm T}\, \,x \Phi_1^{(0\,0)}(x)
       - \frac{p_{+\,\mu}^{\mathrm T} p_{+\,\nu}^{\mathrm T}}{Q^2} 
               x^3\,   \Phi_1^{(0\,0)}. 
        \nonumber 
\end{align}
The leading $t$-dependence is contained in the first structure 
$ g_{\mu\nu}^{\mathrm T}\,$ of (\ref{Wa1})
\begin{align}
\frac{1}{2\pi}{\sf Im}\, T^\twz_{\{\mu\nu\}}\kln{q}|_t = &  g_{\mu\nu}^{\mathrm T}\, 
 \frac{x^2 t}{Q^2}\, \chi(x),\\
 \chi(x)\approx &\bigg\{ \frac{1}{2}x \big(  \Phi_1^{(0\,0)}-  \Phi_1^{(0\,2)}
                 -\eta(3 \Phi_1^{(0\,1)}
                 + \Phi_1^{(0\,3)})\big) -\frac{1}{2}\big( \Phi_1^{(1\,0)} -  
\Phi_1^{(1\,2)}
                 -\eta(3 \Phi_1^{(1\,1)}+ \Phi_1^{(1\,3)})\big)\nonumber \\
         & +\frac{1}{4}x^2\partial_x( \Phi_1^{(0\,0)}- \Phi_1^{(0\,2)}-2\eta  \Phi_1^{(0\,1)})
    -\int_x^1\frac{dy}{y}(\eta \Phi_1^{(1\,1)}
+ \eta \Phi_1^{(1\,3)}+\Phi_1^{(1\,2)})\bigg\}~.
\label{tgsym}
\end{align}
Terms $\propto \eta^2 \simeq x_\PP^2$ are dropped. 

A last remark concerns the generalized Callan-Gross relation (\ref{w1}). 
This 
relation can be written for $\zeta$--moments (\ref{mg}) as
\begin{eqnarray}
{W}^{(n)\rm diff }_{a\,\rm L}(x,p_+^2/Q^2)
&=&
- \,{W}^{(n)\rm diff }_{a\,1}(x,p_+^2/Q^2)
+ \frac{p_+^2 + (qp_+)^2/Q^2}{M^2}\,
 {W}^{(n)\rm diff }_{a\,2}(x,p_+^2/Q^2)\, \nonumber \\
&+&   \frac{p_+^2}{M^2}\int d\zeta \zeta^n \, \epsilon \,  
   {W}^{\rm diff }_{a\,2}(x,{\cal P}^2/Q^2,\zeta ). 
\label{cga} 
\end{eqnarray}
Finally we remark that an equivalent kinematic parameterization can be obtained
using
\begin{eqnarray}
 p_\pm^{\mathrm T} = p_2^{\mathrm T} \pm p_1^{\mathrm T}, 
 \qquad \pi_- = p_- -\eta p_+ = p_2 (1-\eta) - p_1(1+\eta)
       =  p_2^{\mathrm T} (1-\eta) - p_1^{\mathrm T}(1+\eta) .
\end{eqnarray}
These relations allow to link different representations of the hadronic tensor,
which linearly relates various definitions of structure functions, cf.~(\ref{eqD2}).
All contributions due to $M^2$ and $t$--effects in the above are suppressed
like $\propto x^2_{(\PP)} \mu^2/Q^2$ with $\mu^2 = |t|, M^2$.

There are, however also other contributions emerging in the scattering cross 
sections, which are of kinematic origin and stem from 4-vector products 
contributing to the process contracting the leptonic and hadronic tensor, see 
Appendix~A for details. Most of these 
invariants are large, like $l_1.l_2$ and $l_1.p_1$. The invariant 
$l_1.p_2$, (\ref{eqR1}), 
leads to kinematic power corrections further to those considered in 
(\ref{xs1}--\ref{xs3}). Here the leading contribution beyond the lowest
order term is of ${\cal O}(\cos(\phi_b) x_{\rm BJ} \sqrt{\mu^2/Q^2}), \mu^2 = |t|, 
M^2$~.
The terms, which do not vary with the angle $\phi_b$ are of ${\cal O}(x_{\rm BJ} 
\mu^2/Q^2)$. In conclusion, the largest dependences from the limiting case
$|t|, M^2 \rightarrow 0$ are obtained from the kinematic terms in the cross 
section. Those resulting from the target-mass and $t$-corrections of the
hadronic matrix elements always occur with an extra power in $x_{\rm BJ}$ or
$x_\PP$.  

\subsection{The Antisymmetric Part}

\vspace{1mm}\noindent
The contribution to the antisymmetric part of the hadronic tensor is given 
by, cf. \cite{BGR2006,gey2},
\begin{align}
W^\twz_{[\mu\nu]}\kln{q}  &=
 -\pi \,\epsilon_{\mu\nu}^{\phantom{\mu\nu}\alpha\beta} \int d\zeta  \,
   \bigg\{
    \frac{q_{\alpha}\,{\cal K}^a_{5\beta}}{q{\cal P}}
     \Big[g_{a1}(x;\zeta) + g_{a2}(x;\zeta)\Big]
    - \frac{q_{\alpha}\,\mathcal{P}_\beta}{q{\cal P}}
      \frac{(q{\cal K}_5^a)}{q{\cal P}}\; g_{a2}(x;\zeta)
    \nonumber\\
   & \qquad\qquad\qquad\qquad
    + \frac{1}{2}
     \frac{q_{\alpha}\,\mathcal{P}_\beta}{q{\cal P}}
     \frac{(\mathcal{P}{\cal K}_5^a)}{Q^2}\; g_{a0}(x;\zeta)
    \bigg\}\,,
    \label{Tas7}
\end{align}
in terms of the $\zeta$-integral.
Here the coefficients ${\cal K}_{5\,\mu}^a$ are given by (\ref{kinasym+}) and
the functions $g_{a\,k}(x;\zeta)\equiv 
g_{a\,k}(x,\xi,\mathcal{P}^2\!/Q^2,\zeta)|_{k=0,1,2}$ 
read
\begin{align}
   g_{a1}(x;\zeta)&
     =
    \frac{x}{\xi}\frac{1}{[1+x^2\,\mathcal{P}^2/Q^2]^{3/2}}\times
\label{ga1}
    \\
     \qquad\qquad &\quad
    \left[
    \Phi_{5a}^{(0)}(\xi,\zeta)
    +
    \frac{x(\xi +x)\, \mathcal{P}^2/Q^2}{[1+x^2\mathcal{P}^2/Q^2]^{1/2}}
    \Phi_{5a}^{(1)}(\xi,\zeta)
    -
    \frac{x\xi\, \mathcal{P}^2}{2\,Q^2}\, 
    \frac{2-x^2\mathcal{P}^2/Q^2}{1+x^2\mathcal{P}^2/Q^2}
    \Phi_{5a}^{(2)}(\xi,\zeta)
    \right]\!,
    \nonumber\\
      g_{a2}(x;\zeta)
   & =
    - \frac{x}{\xi}\frac{1}{[1+x^2\mathcal{P}^2/Q^2]^{3/2}}\times
\label{ga2}
    \\
    \qquad\qquad&\quad
    \left[
    \Phi_{5a}^{(0)}(\xi,\zeta)
    -
    \frac{1 - x \xi\, \mathcal{P}^2/Q^2}{[1+x^2\mathcal{P}^2/Q^2]^{1/2}}
    \Phi_{5a}^{(1)}(\xi,\zeta)
    -
    \frac{3}{2}
    \frac{x \xi\, \mathcal{P}^2/Q^2}{1+x^2\mathcal{P}^2/Q^2}
    \Phi_{5a}^{(2)}(\xi,\zeta)
    \right]\!,
    \nonumber\\
\intertext{}
  g_{a1}(x;\zeta)& + g_{a2}(x;\zeta) 
   \nonumber \\
    &
= \frac{x}{\xi}\frac{1}{[1+x^2\mathcal{P}^2/Q^2]^{3/2}}
    \left[\left(1 + \frac{x\xi \mathcal{P}^2}{2Q^2}\right)
    \Phi_{5a}^{(1)}(\xi,\zeta)\,
    +\frac{x\xi \mathcal{P}^2}{2Q^2} 
    \Phi_{5a}^{(2)}(\xi,\zeta)\,
    \right]
\label{ga12}
    \\
 &= \frac{x}{\xi}
 		\frac{1}{1+x^2\mathcal{P}^2/Q^2}
 		\left[\Phi_{5a}^{(1)}(\xi,\zeta)\,
     + \frac{1}{2}\,
       \frac{x\xi\, \mathcal{P}^2/Q^2}{[1+x^2\mathcal{P}^2/Q^2]^{1/2}}
    \Phi_{5a}^{(2)}(\xi,\zeta)\right]\,,
    \nonumber\\
   g_{a0}(x;\zeta)
     & 
    = \frac{x^2}{[1+x^2\mathcal{P}^2/Q^2]^{3/2}}
    \times    
\label{ga0} 
    \\     
    &\quad
    \left[
    \Phi_{5a}^{(0)}(\xi,\zeta)
    -
    \frac{3}{[1+x^2\mathcal{P}^2/Q^2]^{1/2}}
    \Phi_{5a}^{(1)}(\xi,\zeta)
    +
    \frac{2- x^2 \mathcal{P}^2/Q^2}{1+x^2\mathcal{P}^2/Q^2}
    \Phi_{5a}^{(2)}(\xi,\zeta)
    \right]\!~.
    \nonumber
\end{align}
The dGPDs $\Phi^{(i)}_{5\,a}(\xi,\zeta)$  are based on (\ref{non15}) 
and, similar to the definitions (\ref{phia0}--\ref{eqphiaa}) 
of $\Phi^{(i)}_{a}(\xi,\zeta)$, 
\begin{eqnarray}
    \Phi^{(0)}_{5\,a}(\xi,\zeta) 
    &\equiv&  \xi\, f_{5\,a}(\xi,\zeta) \,,
\label{phia50}\\
\Phi^{(i)}_{5\,a}\left(\xi,\zeta\right)
    &=& 
    \int^{1}_{\xi}\frac{dy}{y} \,
    \Phi^{(i-1)}_{5\,a}\left(y,\zeta\right)\,, i \geq 1~.
\label{NVF}
\end{eqnarray}
As shown before \cite{BGR2006,gey2} the Wandzura--Wilczek (WW) relation 
\cite{WW} 
holds for the un-integrated distribution functions
\begin{eqnarray}
\label{ww11}
    g^{\twz}_{a2}(x;\zeta)
    &=&
    -\, g_{a1}^{\rm tw2}(x;\zeta) +
    \int_{x}^1 \frac{dy}{y}g_{a1}^{\rm tw2}(y;\zeta)\,,
\end{eqnarray}
between $g_{a2}^{\rm tw2}$ and $g_{a1}^{\rm tw2}$. All target mass and
$t$--corrections can uniquely be
absorbed into the structure functions. Note that this relation holds for
all invariants ${\cal K}_5^a $ independently. The validity of the Wandzura--Wilczek
relation for diffractive scattering at general hadronic scales
$M^2, t$ is a further example in a long list of cases. It was observed using 
the covariant parton model and light-cone expansion
\cite{Roberts:1996ub,Blumlein:1996vs}. For forward scattering, 
target- and quark-mass corrections could be completely absorbed into the 
structure functions 
maintaining the WW-relation \cite{Piccione:1997zh,Blumlein:1998nv}. It is valid
for gluon-induced heavy flavor production \cite{Blumlein:2003wk}, 
non--forward scattering \cite{Blumlein:2000cx}, and
diffractive scattering in the limit $M^2, t \rightarrow 0$
\cite{Blumlein:2002fw}. In the electro--weak case further sum--rules 
exist~\cite{Blumlein:1996vs}. Considering the target mass corrections 
there are  
new twist--3 integral relations~\cite{Blumlein:1998nv}.
The distribution function $g_{a0}^{\rm tw2}$ is also related to $g_{a1}^{\rm tw2}$ but in a more complicated manner:
\begin{eqnarray}
    \label{g02}
    g_{a0}^{\twz}(x;\zeta)
    &=&
    x\xi\,g_{a1}^{\rm tw2}(x;\zeta)
    - \frac{2 x^2 +x\xi}{[1+x^2\mathcal{P}^2/Q^2]^{1/2}}
    \int_{x}^1\!
    \frac{dy}{y}g_{a1}^{\rm tw2}(y;\zeta)
    \nonumber \\
    & &
    +\frac{2x^2}{[1+x^2\mathcal{P}^2/Q^2]^{3/2}}
    \int_{x}^1\! \frac{dy}{y} \!
    \int_y^1\! \frac{dy'}{y'}
    g_{a1}^{\rm tw2}(y';\zeta)\,.
\end{eqnarray}

From (\ref{Tas7}) -- (\ref{ga0}) we now extract the 
$\zeta$-independent functions.
In the kinematic  factors $\zeta$ appears only up to second power.
As preliminary classification we therefore can perform the $\zeta$--integrals according 
to the $\zeta$--powers of the kinematic factors, not counting the internal 
$\zeta$-dependence of the GPDs $ g_{ak}(x;\zeta) $ itself.

For each invariant $a$ let us define
\begin{eqnarray}
\label{eqGG}
    G_{a\,k}^{(n)}(x,\eta,t,p_+^2/Q^2) = 
                \int d\zeta \, \zeta^n \,g_{a\,k}(x;\zeta)\,,
                \qquad
    k = 0,1,2,  
\end{eqnarray}
so that from (\ref{Tas7}) one obtains
\begin{align}
 {\mathrm{Im}}\, T^\twz_{[\mu\nu]}\kln{q}
 &= -\pi  \,\epsilon_{\mu\nu}^{\phantom{\mu\nu}\alpha\beta}\,
   \Bigg\{ 
    \frac{q_{\alpha}\,{\cal K}^a_{5\beta}}{qp_+}
          \Big(G_{a1}^{(0)} + G_{a2}^{(0)}\Big)
    \nonumber\\
   & \quad 
    - \frac{q_{\alpha}\,p_{+\beta}}{qp_+}\left(
              \frac{q{\cal K}_5^a}{qp_+}\, G_{a2}^{(0)}
   - \frac{1}{2}\frac{p_+{\cal K}_5^a}{Q^2}\, G_{a0}^{(0)}
    - \frac{1}{2}
        \frac{\pi_- {\cal K}_5^a}{Q^2}\, G_{a0}^{(1)}
   \right)
    \nonumber\\
    & \quad 
    - \frac{q_{\alpha}\,\pi_{-\beta}}{qp_+}\left(
              \frac{q{\cal K}_5^a}{qp_+}\, G_{a2}^{(1)}
    - \frac{1}{2}\frac{p_+{\cal K}_5^a}{Q^2}\, G_{a0}^{(1)}
    - \frac{1}{2}\frac{\pi_-{\cal K}_5^a}{Q^2}\, G_{a0}^{(2)}\right)\!\!
     \Bigg\}\,.
      \label{as88}
      \\ 
  & \approx
     -\pi  \,\epsilon_{\mu\nu}^{\phantom{\mu\nu}\alpha\beta}\,
   \Bigg\{ 
    \frac{q_{\alpha}\,{\cal K}^a_{5\beta}}{qp_+}
          \Big(G_{a1}^{(0)} + G_{a2}^{(0)}\Big)
- \frac{q_{\alpha}\,p_{+\beta}}{qp_+}
              \frac{q{\cal K}_5^a}{qp_+}\, G_{a2}^{(0)}
    - \frac{q_{\alpha}\,\pi_{-\beta}}{qp_+}
              \frac{q{\cal K}_5^a}{qp_+}\, G_{a2}^{(1)}
\Bigg\}\,. 
\end{align}
In the last line the leading terms are written only.
Now,  inserting the three invariants ${\cal K}_{5}^a $ in (\ref{kinasym+}) 
explicitly, one obtains disregarding sub-leading terms in $1/Q^2$~:
\begin{align}
{\mathrm{Im}}\, &{ T}^{(0)\,\twz}_{[\mu\nu]}\kln{q}
    \approx
    -\pi \, \epsilon_{\mu\nu}^{\phantom{\mu\nu}\alpha\beta}\,
    \Bigg\{
    \frac{q_{\alpha}S^{\mathrm T}_{\beta}}{qp_+}
          \big(G_{1\,1}^{(0)} + G_{1\,2}^{(0)}\big)  
    - \frac{q_{\alpha} p^{\mathrm T}_{+\beta}}{qp_+}
     \bigg[ \frac{q S}{q p_+}\;  G_{1\,2}^{(0)}
    -  
      \frac{p_2.S}{M^2}\;  G_{2\,1}^{(0)} \bigg] 
      \nonumber    \\
&\quad  
  	- \frac{q_{\alpha} \pi^{\mathrm T}_{-\beta}}{qp_+}
   \bigg[ \frac{q S}{q p_+}\;  G_{1\,2}^{(1)}
    + \frac{p_2.S}{M^2} \bigg( G_{2\,2}^{(1)} - G_{3\,1}^{(0)} - 
G_{3\,2}^{(0)}
   \bigg) \bigg] 
    \!\Bigg\}~. 
     \label{ergas0} 
\end{align}
Due to the presence of the Levi-Civita symbol, only transversal 
components (\ref{trans})
contribute. Note that $p_1.S=0$.
The approximate expressions for $ G_{a\,k}^{(n)}$ are
\begin{align}
G_{a1}^{(n)}(x)
     &  \approx  \Phi_{5a}^{(0\,n)}(x) - \frac{x^2\,p_+^2}{Q^2}\gamma_{a1}^n
        \nonumber\\
G_{a2}^{(n)}(x) 
    & \approx
    - \left[
    \Phi_{5a}^{(0\,n)}(x)   - \Phi_{5a}^{(1\,n)}(x) \right]  - 
    \frac{x^2\,p_+^2}{Q^2}\gamma_{a2}^n ,
    \nonumber\\
G_{a1}^{(n)}(x) + G_{a2}^{(n)}(x)
   & \approx    	\Phi_{5a}^{(1\,n)}(x)\, - \frac{x^2\,p_+^2}{Q^2}\gamma_{a12}^n
    \nonumber \\
G_{a0}^{(n)}(x)
    &\approx x^2 \left[
    \Phi_{5a}^{(0\,n)}(x)- {3} \Phi_{5a}^{(1\,n)}(x)
    + {2} \Phi_{5a}^{(2\,n)}(x) \right] -  
\frac{x^2\,p_+^2}{Q^2}\gamma_{a0}^n~,
    \label{G_rel}
\end{align}
with
\begin{align}
\gamma_{a1}^n & = \int d\zeta (1+\epsilon(\zeta)) \zeta^n    
          \left(\frac{5}{4}  \Phi^{(0)}_{a5}(x, \zeta) - 2 \Phi^{(1)}_{a5}(x, \zeta)
           + \Phi^{(2)}_{a5}(x, \zeta)
          +\frac{1}{4}x \partial_{x}\Phi^{(0)}_{a5}(x, \zeta)\right),
\nonumber \\
\gamma_{a2}^n & = \int d\zeta (1+\epsilon(\zeta)) \zeta^n    
          \left(-\frac{5}{4} \Phi^{(0)}_{a5}(x, \zeta) + \frac{11}{4} \Phi^{(1)}_{a5}(x, \zeta)
          -\frac{3}{2}  \Phi^{(2)}_{a5}(x, \zeta)
\right.
\nonumber\\ & \hspace*{6cm} \left.
          -\frac{1}{4} x \partial_{x}\big(\Phi^{(0}_{a5}(x, \zeta)- \Phi^{(1)}_{a5}(x, \zeta)\big)
          \right),
\nonumber \\
\gamma_{a12}^n & = \int d\zeta (1+\epsilon(\zeta)) \zeta^n    
          \left(-\f12 x \Phi^{(2)}_{a5}(x, \zeta) + 
          \frac{3}{4} \Phi^{(1)}_{a5}(x, \zeta)
          +\frac{1}{4} x\partial_x\Phi^{(1)}_{a5}(x, \zeta)\right)
\nonumber 
\intertext{and}
\widetilde\gamma_{a1}^{\,n} & = \int d\zeta  \zeta^n    
          \left(\frac{5}{4}  \Phi^{(0)}_{a5}(x, \zeta) 
- 2 \Phi^{(1)}_{a5}(x, \zeta)
           + \Phi^{(2)}_{a5}(x, \zeta)
          +\frac{1}{4}x \partial_{x}\Phi^{(0)}_{a5}(x, \zeta)\right)~,
\nonumber 
\end{align}
with similar expressions for the other terms $\widetilde\gamma_{ak}^{\,n}$.
The functions $\Phi_{5a}^{(1\,n)}(x)$ are determined as in Eq. (\ref{mph}).
The last function $ G_{a0}^{(n)} $ does not contribute to leading order. As 
final approximate result in leading order  we obtain
\begin{align}
{\mathrm{Im}}\, &{ T}^{(0)\,\twz}_{[\mu\nu]}\kln{q}|_0
    \approx
    -\pi \, \epsilon_{\mu\nu}^{\phantom{\mu\nu}\alpha\beta}\,
    \Bigg\{
    \frac{q_{\alpha}S^{\mathrm T}_{\beta}}{qp_+} \Phi_{51}^{(1,\,0)}(x)
 \label{ergas00}
     \\
  &  + \frac{q_{\alpha} p^{\mathrm T}_{+\beta}}{qp_+} 
     \bigg[ \frac{q S}{q p_+}\;
       \left(\Phi_{51}^{(0,\,0)}(x)-\Phi_{51}^{(1,\,0)}(x)\right)  
    +\frac{p_2 S}{M^2}\Phi_{52}^{(0,\,0)}(x) \bigg]
     \nonumber \\
&  + \frac{q_{\alpha} \pi^{\mathrm T}_{-\beta}}{qp_+}
   \bigg[ \frac{q S}{q p_+}\;\left(\Phi_{51}^{(0,\,1)}(x) 
                         - \Phi_{51}^{(1,\,1)}(x)\right)
   + \frac{p_2 S}{M^2}\left(\Phi_{53}^{(1,\,0)}(x)
   +\Phi_{52}^{(0,\,1)}(x)-  \Phi_{52}^{(1,\,1)}(x)
    \right) \bigg] \Bigg\}~.
  \nonumber  
\end{align}
In Ref.~\cite{Blumlein:2002fw} the terms $ \propto p_2.S$ were neglected 
treating $p_2 || p_1$ and vanishing contributions $\propto \pi_-$. While this 
is correct for $t \rightarrow 0$, a finite 
contribution remains for $M^2 \rightarrow 0$, Eq.~(\ref{eqSp2}),
\begin{eqnarray}
\lim_{t \rightarrow 0} \frac{p_2.S}{M^2} = x_\PP \frac{1-x_\PP/2}{1-x_\PP} 
+ {\cal O}\left(x_{\rm BJ} x_\PP \frac{M^2}{Q^2}\right)~.
\end{eqnarray}
So the previous result \cite{Blumlein:2002fw} is to be modified by a third term
\begin{align}
\label{polNEW}
{\mathrm{Im}}\, { T}^{(0)\,\twz}_{[\mu\nu]}\kln{q}|_0
    \approx
    -\pi \, \epsilon_{\mu\nu}^{\phantom{\mu\nu}\alpha\beta}\,
    \Bigg\{
    \frac{q_{\alpha}S^{\mathrm T}_{\beta}}{qp_+} \Phi_{51}^{(1,\,0)}(x)
  + \frac{q_{\alpha} p^{\mathrm T}_{+\beta}}{qp_+} 
\Bigl[      
\frac{q S}{q p_+}\; 
       \left(\Phi_{51}^{(0,\,0)}(x)-\Phi_{51}^{(1,\,0)}(x)\right)  
      + x_\PP \Phi_{52}^{(0,\,0)}(x) \Bigr]\Bigg\}~.  
\end{align}

The $t$-dependent correction terms result from the $(\eta^2 + t/p_+^2)$--contributions 
in $\epsilon$ and they are entirely 
contained in correction terms $\gamma_{ak}^n $~:
\begin{eqnarray}
\label{ERG}
{\sf Im}\, { T}^{(0)\,\twz}_{[\mu\nu]}\kln{q}|_t
    &\approx&
    +\pi \, \epsilon_{\mu\nu}^{\phantom{\mu\nu}\alpha\beta}\,\frac{x^2\,t}{Q^2}
    \Bigg\{
    \frac{q_{\alpha}S^{\mathrm T}_{\beta}}{qp_+}
          \big( (\widetilde\gamma_{1\,1}^{2} + \widetilde\gamma_{1\,2}^{2})
             +2\eta  (\widetilde\gamma_{1\,1}^{1} + \widetilde\gamma_{1\,2}^{1}) 
             - (\widetilde\gamma_{1\,1}^{0} + \widetilde\gamma_{1\,2}^{0})
                      \big)  \nonumber\\
&&\quad    - \frac{q_{\alpha} p^{\mathrm T}_{+\beta}}{qp_+}
     \bigg[ \frac{q S}{q p_+}\;  
 \big( \widetilde\gamma_{1\,2}^{2} +2\eta \widetilde\gamma_{1\,2}^{1}
           - \widetilde\gamma_{1\,2}^{0}\big) 
  - \frac{p_2 S}{M^2}\; \big( \widetilde\gamma_{2\,1}^{2}
       +2\eta \widetilde\gamma_{2\,1}^{1} -\widetilde\gamma_{2\,1}^{0} 
      \big) \bigg] 
      \nonumber    \\
&&\quad    	- \frac{q_{\alpha} \pi^{\mathrm T}_{-\beta}}{qp_+}
   \bigg[ \frac{q S}{q p_+}\; 
    \big(  \widetilde\gamma_{1\,2}^{3}  +2\eta \widetilde\gamma_{1\,2}^{2}
           - \widetilde\gamma_{1\,2}^{0} \big) \nonumber \\
&&\quad \qquad \qquad    + \frac{p_2 S}{M^2} \bigg(( \widetilde\gamma_{2\,2}^{3} - 
     \widetilde\gamma_{3\,1}^{2} - \widetilde\gamma_{3\,2}^{2})
     +2\eta (\widetilde\gamma_{2\,2}^{2} - 
     \widetilde\gamma_{3\,1}^{1} - \widetilde\gamma_{3\,2}^{1})
\nonumber\\ &&\hspace*{3cm}  
  -(\widetilde\gamma_{2\,2}^{1} - 
     \widetilde\gamma_{3\,1}^{0} - \widetilde\gamma_{3\,2}^{0})
     \bigg) \bigg] 
    \!\Bigg\}~. 
\end{eqnarray}
The corresponding terms are of the same size as in the unpolarized case, Section~4.1, 
and may have a quantitative effect only in the low $Q^2$--region in the meson--exchange case.

It is remarkable that the Wandzura-Wilczek relation \cite{WW} remains intact after 
$\zeta$-integrations and is  valid for the experimentally observable moments,
\begin{eqnarray}
    G^{(n)}_{a\,2}(x,\eta, t, p_+^2/Q^2)  = -\, G_{a\,1}^{(n)}(x,\eta, t, p_+^2/Q^2) 
               + \int_{x}^1\!\! \frac{dy}{y}\;G_{a\,1}^{(n)}(y,\eta, t, p_+^2/Q^2)\,.
    \label{G22}
\end{eqnarray}
The second integral relation (\ref{g02}) contains the $\zeta$--dependent denominator
$\sqrt{1 +x^2 {\cal P}^2/q^2}$ so that we obtain after $\zeta$-integration more
complicated expressions. 
In the approximation  $\pi_- =0 $
one obtains  
\begin{align}
   \label{G02}
    G_{a\,0}^{(n)}(x,\eta, t, p_+^2/Q^2)
    & \approx 
     x\xi_0\;G_{a\,1}^{(n)}(x,\eta, t, p_+^2/Q^2)
    - \frac{2x^2+ x\xi_0}{[1+4x^2\,p_+^2/Q^2]^{1/2}}
    \int_{x}^1\!
    \frac{dy}{y}\;G_{a\,1}^{(n)}(y,\eta, t, p_+^2/Q^2)
    \nonumber \\
    & \quad
    +\frac{2x^2}{[1+4 x^2\, p_+^2/Q^2]^{3/2}}
    \int_{x}^1\!\! \frac{dy}{y} \!
    \int_y^1\! \frac{dy'}{y'}\;
    G_{a\,1}^{(n)}(y',x,\eta, t, p_+^2/Q^2)\,.
\end{align}
$\xi_0$ denotes the Nachtmann variable (\ref{zeta0}).
However the functions $ G_{a\,0}^{(n)}(x,\eta, t, p_+^2/Q^2) $ contribute  to sub-leading terms
only.

\section{Conclusions}
\renewcommand{\theequation}{\thesection.\arabic{equation}}
\setcounter{equation}{0}
\vspace{2mm}
\noindent
Deep--inelastic diffractive scattering, like other hard scattering
processes off nucleons, requires target mass corrections in the region of
lower $Q^2$-scales. In fact, the nucleon mass $M$ is not the only hadronic scale 
relevant to that process where both the incoming and outgoing nucleon play a role. 
The invariant $t = (p_2 - p_1)^2$ on average is of the same size as $M^2$.\footnote{
In case of related semi--exclusive processes in which more than one final--state 
hadron is well separated in rapidity from the inclusively produced hadrons other 
invariants more would emerge.} In the present paper we investigated in detail 
the conditions under which terms like $M^2/Q^2$ or $|t|/Q^2$ contribute.

We considered the leading twist contributions for which factorization theorems
allow a partonic description. With the help of A. Mueller's generalized optical 
theorem it was possible to reformulate diffractive scattering in terms of 
deep--inelastic scattering off an effective two--nucleon pseudo-state 
accounting for $t$. All essential expressions determining experimentally 
relevant quantities are the diffractive generalized parton densities (dGPD) 
defined as expectation values of non-local light-cone operators 
(\ref{non1},\ref{non15}). The involved iterated diffractive dGPDs 
(\ref{phia0}--\ref{eqphiaa2}), respectively (\ref{phia50}) and 
(\ref{NVF}), $ \Phi_{(5)a}^{(i)}(\lambda , \zeta, t , \eta; \mu^2)$ depend on at 
least three variables, $ \lambda , \zeta$ and $ t $. Hereby $t$ is 
an external variable, whereas  $\lambda$ is defined as overall scale 
multiplied with  a generalized momentum in  $(p_+z_+ + p_-z_-) =\lambda {\cal P}$.
In the hadronic tensor $ W_{\mu\nu} = \textrm{Im} \, T_{\mu \nu}$ it 
is fixed by $\xi$, the  generalized Nachtmann variable (\ref{xicd}). Moreover 
the generalized momentum ${\cal P}= p_+ + \zeta \pi_- $ splits into a 
``longitudinal'' and a ``transversal'' part $\pi_-$ multiplied by a new  
variable $\zeta$ and can be treated separately.
The problem in applying the results of our previous work \cite{BGR2006} is the 
dependence of the dGPDs on the `internal' variable $\zeta $ which is not measurable in experiment 
since it contributes through a definite integral in the final 
expressions. We performed an expansion w.r.t. the external variable $p_+^2/Q^2$.
This leads to a set of integrated dGPDs which describe the process and the relevant mass 
corrections in a well-defined approximation. 

One of our results is a prescription of experimental data in terms of 
experimentally accessible integrated diffractive GPD's,
\begin{eqnarray}
\Phi_{(5)a}^{(i\, n)} (\xi, t, \eta)
= \int d\zeta\, \zeta^n \,\Phi_{(5)a}^{(i)} (\xi, \zeta, t,\eta)\,,
\end{eqnarray}
or approximately by the functions (\ref{mph}), 
which could be considered as diffractive parton densities, as it is the case for
vanishing masses \cite{Blumlein:2001xf}. For our approximation a similar relation 
holds,
where $\xi$ is substituted by the variable $x=Q^2/qp_+ $.
Note that one and the same diffractive input GPD 
$\Phi_{(5)a}^{(i)} (\xi, \zeta, t,\eta)$  determines several amplitudes with different 
kinematic factors. This can be seen in  the lowest approximations (\ref{T_01}) or
(\ref{ergas00}) and  for the $t$-dependent corrections  (\ref{tgsym}) and 
(\ref{ERG}). 

The $t$-- and $M^2$--dependence due to the functions $\Phi_{(5)a}^{(in)} 
(\xi, t,\eta)$, besides the non-perturbative $t$--behaviour, turns out to be of
${\cal O}(x^2_{{\rm BJ}(\PP)} \mu^2/Q^2),~\mu^2 = |t|, M^2$. Some of the kinematic factors emerging in 
the 
scattering cross section turn out to be less suppressed and are of ${\cal O}(x_{{\rm BJ}(\PP)} \mu^2/Q^2)$. 
In the case of diffractive scattering the region of $x_{\rm BJ}$ and $x_\PP$ is effectively 
limited 
by $\lessapprox 10^{-2}$. The corresponding corrections cannot be resolved at the 
experimental 
accuracy. The effects are larger in the case of meson--exchange processes with a fast hadron 
due to the range $x \lessapprox 0.3$.
Due to the smallness of these corrections the diffractive distribution 
functions obey a partonic description, where $t$ plays the role of an additional variable 
besides $\beta=x_{\rm BJ}/x_\PP$.

At the level of twist--2 the structure functions the scattering cross section can be built from 
the corresponding operator expectation values (\ref{non1}--\ref{non15}) as in the case of
deep--inelastic scattering since the specifics of diffractive scattering is
moved into the corresponding two--particle wave functions. Consequently,
the logarithmic scaling violations, which can be completely associated
with that of the {\sf operators},
cf.~\cite{Blumlein:1999sc,Blumlein:2001xf}, are found to be the same as in
DIS or DVCS, if the complete diffractive GPDs are used.

The integral relations (\ref{w1}), (\ref{ww11}) and (\ref{g02}) can be transformed in part 
to the integrated functions only.
The presence of target mass and $t$-effects enlarges the number of
structure functions determining the hadronic tensor if compared to the case of forward scattering.
As shown in the present paper, these corrections are suppressed by at least one power in $x_{\rm 
BJ}$ or $x_\PP$ and therefore the picture derived in \cite{Blumlein:2001xf,Blumlein:2002fw} remains 
valid quantitatively. In the polarized case, there is a new term, cf. (\ref{polNEW}), which 
contains $x_\PP$ as prefactor. The Wandzura--Wilzcek relation remains unbroken
and holds even separately for the contributions of the three different invariants
${\cal K}_a^5$ (\ref{kinasym+}). We have also shown how the present formalism can be used
to derive the target mass corrections in the limit of forward scattering.

\appendix
\newpage
\section{Kinematic Relations}
\renewcommand{\theequation}{\thesection.\arabic{equation}}
\setcounter{equation}{0}

\vspace{1mm}
\noindent
In the following we list kinematic relations for the process of deeply-inelastic 
diffractive scattering. The incoming and outgoing lepton momenta are $l_1$ and $l_2$,
those of the nucleon are $p_1$ and $p_2$ (diffractive nucleon), and the vector of the 
remainder hadrons is denoted by $r$. We disregard the lepton masses, $l_1.l_1=l_2.l_2=0$.
The kinematic invariants of this $2 \rightarrow 3$ 
particle scattering process are, cf.~\cite{KB},
\begin{eqnarray}
p_1.p_1 &=& p_2.p_2 = M^2~,                   \\
r.r   &=& M_X^2~,                         \\
s     &=& (l_1 + p_1)^2 = 2 l_1.p_1 +M^2~,  \\
q^2   &=& - Q^2 = (l_1-l_2)^2 = - 2 l_1.l_2~, \\
t     &=& (p_1-p_2)^2 = 2 M^2 - 2 p_1.p_2~,  \\
W^2   &=& (r+p_2)^2 = (q+p_1)^2 = Q^2 \left(\frac{1}{x_{\rm BJ}} -1 \right) + M^2~, \\
l_1.q &=&  - Q^2/2~, \\ 
l_2.q &=&  + Q^2/2~, \\
s_1   &=& (l_2+r)^2~,          \\
2 l_1.p_2 &=& s - s_1 + t- M^2~.
\end{eqnarray}
For the later analysis it will be useful to consider the cms frame of the momenta
\begin{eqnarray}
\label{eqFR1}
\pvec_1 + \qvec = \pvec_2 + \rvec = 0~.
\end{eqnarray}
We need to express $S_{||}.p_2$. This requires a suitable representation of $p_2$,
which cannot be obtained from the invariants above. In the frame~(\ref{eqFR1}) the 
energies and absolute values of the three-momenta are given by 
\begin{eqnarray}
E_q     &=& \frac{1}{2 \sqrt{W^2}}\left[W^2-Q^2-M^2\right]~, \\
E_{p_1} &=& \frac{1}{2 \sqrt{W^2}}\left[W^2+Q^2+M^2\right]~, \\
|\qvec| &=& |\pvec_1| = \frac{1}{2 \sqrt{W^2}}\lambda^{1/2}(W^2,-Q^2,M^2)~,\\
E_r     &=& \frac{1}{2 \sqrt{W^2}}\left[W^2+M_X^2-M_1^2\right]~, \\
E_{p_2} &=& \frac{1}{2 \sqrt{W^2}}\left[W^2+M^2-M_X^2\right]~, \\
|\rvec| &=& |\pvec_2| = \frac{1}{2 \sqrt{W^2}}\lambda^{1/2}(W^2,M^2,M_X^2)~,\\
E_l     &=& |\lvec|   = \frac{1}{2 \sqrt{W^2}} \left[s - Q^2 - M^2\right]~.
\end{eqnarray}
The spin vector $S_{||}$ and the four vector $p_2$ read
\begin{eqnarray}
S_{||} &=& \frac{1}{2\sqrt{W^2}} \left( \lambda^{1/2}(W^2,-Q^2,M^2); 
0,0,W^2+M^2+Q^2\right)~,
\\
p_2 &=& \frac{1}{2\sqrt{W^2}} \left ( W^2 + M^2 -M_X^2; \pvec_{\perp,2}, 
\cos\theta_{1,2} \lambda^{1/2}(W^2, M^2, M_X^2)\right)~,   
\end{eqnarray}
with $S_{||}^2 = - M^2$ and
\begin{eqnarray}
\cos\theta_{1,2} &=& \frac{2 W^2 (t - 2 M^2) + (W^2+Q^2-M^2)(W^2+M^2-M_X^2)}
                   {\sqrt{\lambda(W^2,M^2,-Q^2) \lambda(W^2, M^2, M_X^2)}}\nonumber\\
                 &=&\frac{1 - x_\PP + \frac{tx_{\rm BJ}}{Q^2} 
- \frac{4x_{\rm BJ}M^2}{Q^2} \left(1-x_{\rm BJ}+ x_{\rm BJ}\frac{M^2}{Q^2}\right) 
- 2 \frac{x^2_{\rm BJ}t}{Q^2}\left(1-\frac{M^2}{Q^2}\right)} 
{\left\{\left(1+\frac{4 x^2_{\rm BJ} M^2}{Q^2}\right) \left[\left(1-x_\PP - 
\frac{x_{\rm BJ}t}{Q^2}\right)^2 - 4 x_{\rm BJ} x_\PP \frac{M^2}{Q^2} \left(1- \beta + \beta 
\frac{t}{Q^2}\right)\right]\right\}} \nonumber\\
&\simeq& 1 - \frac{x_{\rm BJ}}{1-x_\PP} \left[\frac{|t|}{Q^2} \left(1 
+\frac{2}{1-x_\PP}\right) + \frac{4 M^2}{Q^2} \right] + 
{\cal O}\left(\left(x^2_{\rm BJ}, \frac{x_{\rm BJ} \mu^2}{Q^2}\right)^2\right)~, 
\end{eqnarray}
with $\mu^2 = t, M^2$. Note that the dependence on $\mu^2/Q^2$ is here 
{\it linear} with $x_{\rm BJ}$. 
\begin{eqnarray}
\lambda(a,b,c) = (a-b-c)^2 -4b c
\end{eqnarray}
denotes the triangle-function.
In the limit $t, M^2 \rightarrow 0$ one obtains $\cos\theta_{1,2} = 1$. 

$S_{||}.p_2$ is given by
\begin{eqnarray}
\label{eqSp2}
S_{||}.p_2 &=& \frac{1}{4 W^2} \Bigl[\lambda^{1/2}(W^2,-Q^2,M^2) 
(W^2+M^2-M_X^2) \nonumber \\
& & \hspace*{2cm}
- \cos(\theta_{1,2}) \lambda^{1/2}(W^2,M^2,M_X^2) (W^2+M^2+Q^2) \Bigr]
\nonumber\\ 
&=& \frac{M^2 x_\PP (1-x_\PP/2)}{1-x_\PP} + \frac{|t| 
(3-x_\PP)}{4(1-x_{\rm BJ})(1-x_\PP)} + {\cal O}(|t|^2, M^4, |t| M^2)~.
\end{eqnarray}

Further $S_{||}.l_1$ and $S_{||}.q$ are
\begin{eqnarray}
S_{||}.l_1 &=& \frac{1}{4 W^2} \left(s - Q^2 - M^2\right) \frac{Q^2}{x_{\rm BJ}} \left[\left(
1 + \frac{4 x^2_{\rm BJ} M^2}{Q^2}\right)^{1/2} - \left(1+ \frac{2x_{\rm BJ}M^2}{Q^2}\right) 
\right]
\nonumber\\
&\simeq& - \frac{1}{2y} (1-x_{\rm BJ}y)  M^2 + {\cal O}\left(\frac{x_{\rm BJ}^2 
M^4}{Q^2} 
\right)~, \\
S_{||}.q &=& - \frac{1}{2 W^2} \left(Q^2 - M^2\right) \frac{Q^2}{x_{\rm BJ}} \sqrt{1 
+ \frac{4 x^2_{\rm BJ} M^2}{Q^2}} \nonumber\\ 
&=& - \frac{Q^2}{1-x_{\rm BJ}} \left[1- \frac{M^2}{Q^2}\left(\frac{1}{1-x_{\rm BJ}} - 4 
x^2_{\rm BJ}\right) 
+ {\cal O}\left(\left(x_{\rm BJ} \frac{M^2}{Q^2}\right)^2\right)\right]~.
\end{eqnarray}
Note that these expressions contain terms of ${\cal O}(M^2/Q^2)$ and ${\cal O}(x_{\rm BJ} M^2/Q^2)$.
$S_{||}.l_1$  and $S_{||}.p_2$ vanish in the strict collinear limit $t, M^2 
\rightarrow 0$.

The above invariants, except $s_1$,  were all parameterized in terms of the dimensionless
quantities, as $x_{\rm BJ}, y, x_\PP$ keeping $M^2$ and $t$, which are normalized to 
$Q^2$. The invariant
\begin{eqnarray}
s_1 = s + M^2 - \frac{1}{\lambda(W^2,q^2,M^2)} \left[D_1 + 2 \cos(\phi_b) 
\sqrt{G_1 
G_2}\right]~, 
\end{eqnarray}
in addition depends on the azimuthal angle $\phi_b$. Here,
\begin{eqnarray}
G_1 &=& G(s,q^2,W^2,0,M^2,0) \leq 0~, \\
G_2 &=& G(W^2,t,M^2,q^2,M^2,M_X^2) \leq 0~,  
\end{eqnarray}
where $G$ denotes the Caley determinant 
\begin{eqnarray}
G(x,y,z,u,v,w) = - \frac{1}{2} \left|
\begin{array}{ccccc}
                 0 & 1 & 1 & 1 & 1 \\
                 1 & 0 & v & x & z \\
                 1 & v & 0 & u & y \\
                 1 & x & u & 0 & w \\
                 1 & z & y & w & 0 
\end{array}\right|~.
\end{eqnarray}
$D_1$ is the determinant
\begin{eqnarray}
D_1 = \left|
\begin{array}{lll}
2 M^2         & W^2-q^2 + M^2 & 2 M^2 - t \\
W^2-q^2 + M^2 & 2 W^2         & W^2-M_X^2 + M^2 \\
s + M^2       & s + W^2       & 0
\end{array}\right|~.
\end{eqnarray}
Let us consider the limit $M^2, t \rightarrow 0$. Here 
\begin{eqnarray}
G_2 = G(W^2,0,0,q^2,0,M_X^2) = 0~,
\end{eqnarray}
and $s_1$ does not depend on the azimuthal angle $\phi_b$. Furthermore, 
\begin{eqnarray}
D_1       &=& (W^2+Q^2)(W^2-M_X^2)s = \frac{s Q^4}{x^2_{\rm BJ}}(1-x_\PP)~,\\
2 l_1.p_2 &=& s(1-x_\PP)~.
\end{eqnarray}
Therefore we obtain in the limit $M^2, t \rightarrow 0$ the hadronic tensors given in
\cite{Blumlein:2001xf,Blumlein:2002fw}.

We now expand $2 l_1.p_2$ up to terms linear in $M^2$ and $t$. One obtains 
\begin{eqnarray} 
G_1 &\simeq& -s^2 Q^2 \left[(1-y) - 
\frac{M^2}{Q^2} x_{\rm BJ} y \left(2-y(1-x)\right)\right]~,\\ 
G_2 &\simeq& - \frac{Q^4}{x_{\rm BJ}^2} \left[ \left(1-2x_{\rm BJ}-x_\PP\right) \frac{|t|}{Q^2} 
- \frac{M^2}{Q^2}  x_\PP^2 \left(1+2 
\beta\right)^2\right]~,
\\
D_1 &\simeq&  \frac{Q^6 }{y x_{\rm
BJ}^3} \Bigl[1-x_\PP+\frac{|t|}{Q^2} x_{\rm BJ} 
(y(1-x_{\rm BJ})+2x_{\rm BJ})
+ 2 \frac{M^2}{Q^2} x_{\rm BJ}(2 x_{\rm BJ} + y(1 -  x_{\rm BJ} x_\PP)) 
\nonumber\\ &&
\hspace*{2.2cm}+ {\cal O}(\mu^4/Q^4) \Bigr]~.
\end{eqnarray} 
The ratio $l_1.p_2/l_1.p_1$ receives $\sqrt{\mu^2/Q^2}$ corrections for the 
angular term $\propto \cos \phi_b$ and $\mu^2/Q^2$ corrections otherwise,
\begin{eqnarray}
\label{eqR1}
\frac{l_1.p_2}{l_1.p_1} &=& 1 - x_\PP + \frac{|t|}{Q^2} x_{\rm BJ} 
\left[y(1-x_{\rm BJ}) + 2 x_{\rm BJ}\right] + 2 \frac{M^2}{Q^2} x_{\rm BJ} y 
(1- x_{\rm BJ} x_\PP) \nonumber\\ & & \hspace*{7mm}
+ 2 \cos\phi_b x \sqrt{1-y} \left[(1-2 x_{\rm BJ} - x_\PP) \frac{|t|}{Q^2}
- \frac{M^2}{Q^2} x_\PP^2(1+2\beta)^2\right]^{1/2} 
\nonumber\\ && + {\cal O}\left(\frac{\mu^3}{(Q^2)^{3/2}}\right)~.
\end{eqnarray}

\newpage
\section{The Limiting Case of Deep--Inelastic Scattering}
\setcounter{equation}{0}
\label{sec-DIS}

\noindent
As a check of our general result we perform the limit
$p_2 \rightarrow 0$ to obtain the results of 
Refs.~\cite{Georgi:1976ve,Piccione:1997zh,Blumlein:1998nv}.
In this limit the kinematic variables and invariants are given by
\begin{eqnarray}
\label{d1}
& &{\cal P} \rightarrow  p_1 \equiv p\,,
\qquad
 x \rightarrow  2 x_{\rm Bj}^f \equiv \frac{Q^2}{qp}\,,
 \qquad
 \eta \rightarrow -1\,,
 \qquad
 \pi_- \rightarrow 0 \,,
  \\
& &{\cal P}^2 \rightarrow M^2\,,
    \quad
    t \rightarrow M^2\,,
    \quad
    p_- p_+ \rightarrow -M^2\,,
 \qquad
 {\cal K }^{1} \rightarrow  p \,,
 \qquad
 {\cal K }^{2} \rightarrow 0 \,.
 \label{d2}
\end{eqnarray}
The generalized Nachtmann variable takes the form
\begin{eqnarray}
\qquad
\xi  ~\rightarrow~
   2\,\frac{2 x_{\rm Bj}^f}{1 + \sqrt{1 + 4x_{\rm Bj}^f M^2/Q^2}}
 = 2 \xi^f\,.
 \label{xid1}
\end{eqnarray}
First, we consider the symmetric part of the amplitude. The second kinematic 
variable  
${\cal K}^2 =\pi_- $ vanishes. In (\ref{TSext}) only the contributions for $a=1$ remain,
\begin{align}
{\sf Im}\, T^\twz_{1\,\{\mu\nu\}}\kln{q}
 &= 2 \pi \int d\zeta\;
    \bigg[
 - g_{\mu\nu}^{\mathrm T}\, {W}^{\rm diff}_{1\,1} (x,{\cal P}^2/Q^2;\zeta)
 + \frac{p_\mu^{\mathrm T} p_\nu^{\mathrm T}}{M^2}
 \,{W}^{\rm diff}_{1\,2}(x,{\cal P}^2/Q^2;\zeta)\bigg]\,
 \label{Ts_enddis}
 \\
&\rightarrow  2 \pi
 \bigg[
 - g_{\mu\nu}^{\mathrm T}\, {W}_1 (\xi^f)
 + \frac{p_\mu^{\mathrm T} p_\nu^{\mathrm T}}{M^2}
 \,{W}_2(\xi^f)\bigg]\,.
\end{align}
Because of $p_2=0$, the integration over $z_2$ can now be performed,
\begin{equation}
\int dz_2\, \phi(z_1,z_2)= \widehat\phi(z_1)~, 
\end{equation}
where $\widehat{\Phi}(z_1)$ denotes the parton density in the deep--inelastic case.
The variables $z_i$ are expressed by
\begin{eqnarray}
  z_1 \rightarrow \lambda = \xi  \, , \quad
  z_2 \rightarrow \lambda ( 2\zeta +1)= \xi ( 2\zeta +1)\,,\quad
  dz_2 = 2 \,\xi \,d\zeta \,.
\end{eqnarray}
From the complete integration measure $2|\lambda|d\lambda d\zeta$ the
$\lambda$-integral has already been carried out, so that only the
$\zeta$-integration remains.

To get the standard structure functions for deep--inelastic scattering
we take the limits (\ref{d1}--\ref{xid1}) and perform the $\zeta$-integration,
\begin{eqnarray}
\label{limdi}
 W_k(\xi^f,x_{\rm Bj}^f ,p^2/Q^2)  
 =\int d\zeta \,\lim_{p_2 \rightarrow 0} {W}^{\rm diff}_{1\,k} (\xi,x,{\cal 
P}^2/Q^2;\zeta)
 \qquad {\rm for} \qquad k = 1,2\,.
\end{eqnarray}
To obtain explicit expressions we use 
${W}^{\rm diff}_{1\,1}(\xi, x, {\cal P}^2/Q^2;\zeta)$ and
${W}^{\rm diff}_{1\,2}(\xi, x, {\cal P}^2/Q^2;\zeta)$ 
in (\ref{WF1}) and (\ref{WFj})
together with the diffractive structure functions $F_{1\,1}(\xi,\zeta)$ and 
$F_{1\,2}(\xi,\zeta)$  
as given by (\ref{F1x}) and (\ref{F2x}), respectively. We obtain
\begin{eqnarray}
W_1  =
  \frac{2\,x_{\rm Bj}^f}{\!\sqrt{1+ 4 (x^{f}_{\rm Bj})^2 M^2/Q^2}}
  \Bigg[
  \widehat\Phi^{(0)}_{f\,1}
  +\frac{x_{\rm Bj}^f{ M^2}/{Q^2}}{\!\sqrt{1+ 4 (x^{f}_{\rm Bj})^2 M^2/Q^2}}\,
  \widehat\Phi^{(1)}_{f\,1}
  +\frac{(x_{\rm Bj}^{f}\,{ M^2}/{Q^2})^2}{1+ 4 (x^{f}_{\rm Bj})^2 M^2/Q^2}\,
  \widehat\Phi^{(2)}_{f\,1}
  \Bigg]
  \nonumber\\
\end{eqnarray}
and
\begin{eqnarray}
W_2  =
  \frac{(2\,x_{\rm Bj}^f)^3 M^2/Q^2}{\!{\sqrt{1+ 4 (x^{f}_{\rm Bj})^2 M^2/Q^2}}^{\,\,3}}
  \Bigg[
  \widehat\Phi^{(0)}_{f\,1}
  +\frac{3\,x_{\rm Bj}^f\,{M^2}/{Q^2}}{\!\sqrt{1+ 4 (x^{f}_{\rm Bj})^2 M^2/Q^2}}\, 
  \widehat\Phi^{(1)}_{f\,1}
  +\frac{3\,(x_{\rm Bj}^f\,{M^2}/{Q^2})^2}{1+ 4 (x^{f}_{\rm Bj})^2 M^2/Q^2}\,
  \widehat\Phi^{(2)}_{f\,1}\!
  \Bigg]~, \nonumber\\ 
\end{eqnarray}
where the functions $\widehat\Phi^{(n)}_{f\,1}(2\xi^f) $ follow from (\ref{phia0}--\ref{phia2}).

The $\zeta$--integrals can be performed taking into account
\begin{eqnarray}
\widehat\Phi^{(n)}_{f\,1}(2\xi^f) = 2^n\,\Phi^{(n)}_{f\,1}(\xi^f), 
\end{eqnarray}
which yields
\begin{align}
\label{phiaf0}
\int d\zeta\Phi_a^{(0)}(\xi,\zeta) &\equiv \widehat\Phi_a^{(0)}(\xi)
\rightarrow
\widehat\Phi_{f\,a}^{(0)}(2\xi^f)
 = f_{f\, a} (\xi^f)\, =\Phi_{f\,a}^{(0)}(\xi^f)\,,
\\
\label{phiaf1}
\int d\zeta\Phi_a^{(1)}(\xi,\zeta)
&\equiv \widehat\Phi_a^{(1)}(\xi)
\rightarrow
\widehat\Phi_{f\,a}^{(1)}(2\xi^f)
  =  2 \int_{\xi^f}^{1} dy_1 \;f_{f\,a} (y_1) =2 \Phi_{f\,a}^{(1)}(\xi^f)\,,
\\
\label{phiaf2}
\int d\zeta\Phi_a^{(2)}(\xi,\zeta)
&\equiv \widehat\Phi_a^{(2)}(\xi)
\rightarrow
\widehat\Phi_{f\,a}^{(2)}(2\xi^f)=
4 \int_{\xi^f}^{1} dy_2
\int_{y_2}^{1} dy_1 \, f_{f\,a} (y_1) = 4\Phi_{f\,a}^{(2)}(\xi^f)\,,
\end{align}
in the limit $p_2 \rightarrow 0$.
Finally one obtains
\begin{align}
W_1 (\xi^f)& =
  \frac{2\,x_{\rm Bj}^f}{\!\sqrt{1+ 4 (x^f_{\rm Bj})^2 M^2/Q^2}}
  \Bigg[
  \Phi^{(0)}_{f\,1}
  +\frac{2\,x_{\rm Bj}^f{ M^2}/{Q^2}}{\!\sqrt{1+ 4 (x^f_{\rm Bj})^2M^2/Q^2}}
   \Phi^{(1)}_{f\,1}
  +\frac{4\,(x_{\rm Bj}^f\,{ M^2}/{Q^2})^2}{{1+ 4 (x^f_{\rm Bj})^2M^2/Q^2}}
    \Phi^{(2)}_{f\,1}
  \Bigg]
\intertext{and}
W_2 (\xi^f) &=
  \frac{(2\,x_{\rm Bj}^f)^3 M^2/Q^2}{\!\sqrt{1+ 4 (x^f_{\rm Bj})^2M^2/Q^2}^{\,\,3}}
  \Bigg[
   \Phi^{(0)}_{f\,1}
  +\frac{6\,x_{\rm Bj}^f{ M^2}/{Q^2}}{\!\sqrt{1+ 4 (x^f_{\rm Bj})^2M^2/Q^2}}
  \Phi^{(1)}_{f\,1} 
  +\frac{12\,(x_{\rm Bj}^f\,{ M^2}/{Q^2})^2}{{1+ 4 (x^f_{\rm Bj})^2M^2/Q^2}}
  \Phi^{(2)}_{f\,1}
  \Bigg]\!~,
  \nonumber\\
\end{align}
the representation for the target mass corrections in the unpolarized case given in 
\cite{Georgi:1976ve} before.

As in the case of generalized parton densities also here the diffractive hadronic 
distribution amplitudes contain as limit the parton
distribution of deeply inelastic scattering. However, care is needed
because
\begin{eqnarray}
  \Phi^{(0)}_{f \,a}(\xi_f) = \widehat \Phi_a(2\xi^f, t=M^2)
\end{eqnarray}
includes an analytic continuation from the physical values $t<0$ to $ t=M^2$.

Next, we consider the antisymmetric contributions in the Compton amplitude, which 
correspond to the case of polarized scattering. From the kinematic factors 
(\ref{kinasym+}) only  ${\cal K}^1_{5} = S$ remains in the limit $p_2 \rightarrow 0$. 
We consider (\ref{as88}) and (\ref{ergas0}) with the definition (\ref{eqGG}) for $ 
G_{1\,k}^{(n)}$. This results in
\begin{eqnarray}
 \label{Tas8d}
 {\mathrm{Im}}\, T^\twz_{[\mu\nu]\,f}\kln{q}
 &=&
 \pi  \,\epsilon_{\mu\nu}^{\phantom{\mu\nu}\alpha\beta}\,
   \bigg\{ 
   \frac{q_{\alpha}S_{\beta}}{qp}\Big(G_{11}^{(0)}(x^f_{\rm Bj}) + 
G_{12}^{(0)}(x^f_{\rm Bj})\Big)
   - \frac{q_{\alpha}p_{\beta}}{qp}\,\frac{qS}{qp}\; G_{12}^{(0)}(x^f_{\rm Bj})
   \bigg\}\,,
\end{eqnarray}
 the forward scattering limit (\ref{d1}) 
of our general result (\ref{as88}) and (\ref{ergas0}) 
with the definition (\ref{eqGG}) of $ G_{1\,k}^{(n)}$.
The antisymmetric part of the amplitude simplifies to 
\begin{align}
\label{G_11}
 G_{11}^{(0)}(x^f_{\rm Bj})&
  =
    \frac{x^f_{\rm Bj}/\xi^f }{[1+4(x^f_{\rm Bj})^2 M^2/Q^2]^{3/2}}
    \bigg[\widehat{\Phi}_{51}^{(0)}(2\xi^f)
    +
    \frac{4x^f_{\rm Bj} (x^f_{\rm Bj} +\xi^f)M^2/Q^2}{[1+(4x^f_{\rm Bj})^2M^2/Q^2]^{1/2}}\;
    \widehat{\Phi}_{51}^{(1)}(2\xi^f)
\nonumber\\
  &  \qquad  \qquad \qquad
    -2 x^f_{\rm Bj} \xi^f M^2/Q^2
    \frac{2-4(x^f_{\rm Bj})^2M^2/Q^2}{1+4(x^f_{\rm Bj})^2 M^2/Q^2}\;
    \widehat{\Phi}_{51}^{(2)}(2\xi^f)
    \bigg],
    \\
\label{G_12}
G_{12}^{(0)}(x^f_{\rm Bj})&
    =
    \frac{-x^f_{\rm Bj}/\xi^f }{[1+4 (x^f_{\rm Bj})^2 M^2/Q^2]^{3/2}}
    \bigg[
    \widehat{\Phi}_{51}^{(0)}(2\xi^f)
    -
    \frac{1 -4x^f_{\rm Bj} \xi^f M^2/Q^2}{[1+4(x^f_{\rm Bj})^2M^2/Q^2]^{1/2}}\;
    \widehat{\Phi}_{51}^{(1)}(2\xi^f)
\nonumber\\
    & \qquad  \qquad\qquad
    - \frac{6\,x^f_{\rm Bj} \xi^f M^2/Q^2}{1+ 4 (x^f_{\rm Bj})^2 M^2/Q^2}\;
    \widehat{\Phi}_{51}^{(2)}(2\xi^f)
    \bigg],
\\
\intertext{and }
\label{G_112}
    G_{11}^{(0)}(x^f_{\rm Bj})& + G_{12}^{(0)}(x^f_{\rm Bj})=
      \frac{{x^f_{\rm Bj}}/{\xi^f}}{[1+4(x^f_{\rm Bj})^2 M^2/Q^2]^{3/2}} \\
    &  \qquad  \qquad \qquad
    \times\,
    \left[\left(1 +2x^f_{\rm Bj} \xi^f M^2/Q^2\right)
    \widehat{\Phi}_{51}^{(1)}(2\xi^f)\,
    +2x^f_{\rm Bj}\xi^f M^2/Q^2\;
    \widehat{\Phi}_{51}^{(2)}(2\xi^f)\,
    \right].
  \nonumber
\end{align}
Again we introduced the (integrated) parton distributions $\widehat\Phi_{5\,a}^{(0)}(\xi)$ 
and performed the limit (\ref{d1}) as follows:
\begin{eqnarray}
\label{phiaf50}
\int d\zeta\Phi_{5\,a}^{(0)}(\xi,\zeta)
&\equiv&
\widehat\Phi_{5\,a}^{(0)}(\xi)
 \rightarrow
 \widehat\Phi_{5\,a}^{(0)}(2\xi^f)=
 2\xi^f f_{5\,f\, a} (\xi^f)\, =\Phi_{5\,f\,a}^{(0)}(\xi^f)\,,
\\
\label{phiaf51}
\int d\zeta\Phi_{5\,a}^{(1)}(\xi,\zeta)
&\equiv&
\widehat\Phi_{5\,a}^{(1)}(\xi)
\rightarrow
\widehat\Phi_{5\,a}^{(1)}(2\xi^f)=
     \int_{\xi^f}^{1}\frac {dy_1}{y_1} \;\Phi_{5\,f\,a}^{(0)}(y_1)
     =  \Phi_{5\,f\,a}^{(1)}(\xi^f)\,,
\\
\label{phiaf52}
\int d\zeta\Phi_{5\,a}^{(2)}(\xi,\zeta)
&\equiv&
\widehat\Phi_{5\,a}^{(2)}(\xi)
\rightarrow
\widehat\Phi_{5\,a}^{(2)}(2\xi^f)=
      \int_{\xi^f}^{1} \frac{dy_2}{y_2} \int_{y_2}^{1}
      \frac{dy_1}{y_1} \,\Phi_{5\,f\,a}^{(0)}(y_1)
     = \Phi_{5\,f\,a}^{(2)}(\xi^f)\,~.
\end{eqnarray}
Finally we substitute $\widehat\Phi_{5\,a}^{(i)}(2\xi^f)$
by $ \Phi_{5\,f\,a}^{(2)}(\xi^f)$ and obtain the result given in 
\cite{Blumlein:1998nv,Piccione:1997zh} before.

\bigskip

\noindent
{\large{\bf Acknowledgment}}\\

\noindent
This paper was supported in part by SFB-TR-9: Computergest\"utze
Theoretische Teilchenphysik.
We thank M. Diehl, P. Kroll, D. M\"uller, W.-D. Nowak
and M. Vanderhaeghen for discussions.

\newpage

\end{document}